\documentclass[aps,twocolumn, prX,nofootinbib]{revtex4}
\usepackage[usenames, dvipsnames]{xcolor}
\usepackage{graphicx, array}
\usepackage{graphics,epsfig}
\usepackage{amssymb}
\usepackage{amsmath}
\usepackage{ulem}
\usepackage{multirow}
\usepackage{subfigure}
\usepackage{ulem}
\usepackage{gensymb}
\usepackage{natbib}
\usepackage{appendix}

\makeatletter
\def\p@subsection{}
\makeatother

\usepackage{listings}
\usepackage{aas_macros}
\usepackage{comment}
\usepackage{soul}

\newcommand{\tauth}{\bar{\tau}_\mathrm{theory}}
\newcommand{\seof}{\textrm{DR5 f150}}
\newcommand{\sen}{\textrm{DR5 f090}}
\newcommand{\ILC}{\textrm{DR4 ILC}}
\newcommand{\sigt}{\sigma_\textrm{T}}
\newcommand{\rmn}{\mathrm}
\usepackage{hyperref}

\begin{document}

\input{epsf}

\title{The Atacama Cosmology Telescope: Probing the baryon content of SDSS DR15 galaxies with the thermal and kinematic Sunyaev-Zel'dovich effects}

\author{E.~M.~Vavagiakis$^1$}
\author{P.~A.~Gallardo$^1$}
\author{V.~Calafut$^2$}
\author{S.~Amodeo$^2$}
\author{S.~Aiola$^{3}$}
\author{J.~E.~Austermann$^4$}
\author{N.~Battaglia$^2$}
\author{E.~S.~Battistelli$^5$}
\author{J.~A.~Beall$^4$}
\author{R.~Bean$^2$}
\author{J.~R.~Bond$^6$}
\author{E.~Calabrese$^7$}
\author{S.~K.~Choi$^{1,2}$}
\author{N.~F.~Cothard$^8$}
\author{M.~J.~Devlin$^9$}
\author{C.~J.~Duell$^1$}
\author{S.~M.~Duff$^4$}
\author{A.~J.~Duivenvoorden$^{10}$}
\author{J.~Dunkley$^{10,11}$}
\author{R.~Dunner$^{12}$}
\author{S.~Ferraro$^{13,14}$}
\author{Y.~Guan$^{15}$}
\author{J.~C.~Hill$^{3,16}$}
\author{G.~C.~Hilton$^4$}
\author{M.~Hilton$^{17}$}
\author{R. Hlo\v{z}ek$^{18,19}$}
\author{Z.~B.~Huber$^1$}
\author{J.~Hubmayr$^4$}
\author{K.~M.~Huffenberger$^{20}$}
\author{J.~P.~Hughes$^{21}$}
\author{B.~J.~Koopman$^{22}$}
\author{A.~Kosowsky$^{15}$}
\author{Y.~Li$^1$}
\author{M.~Lokken$^{6,18,19}$}
\author{M.~Madhavacheril$^{23}$}
\author{J.~McMahon$^{24,25,26,27}$}
\author{K.~Moodley$^{17,28}$}
\author{S.~Naess$^{3}$}
\author{F.~Nati$^{29}$}
\author{L.~B.~Newburgh$^{22}$}
\author{M.~D.~Niemack$^{1,2}$}
\author{L.~A.~Page$^{10}$}
\author{B.~Partridge$^{30}$}
\author{E.~Schaan$^{12,13}$}
\author{A.~Schillaci$^{31}$}
\author{C.~Sif\'on$^{32}$}
\author{D.~N.~Spergel$^{3,11}$}
\author{S.~T.~Staggs$^{10}$}
\author{J.~N.~Ullom$^4$}
\author{L.~R.~Vale$^4$}
\author{A.~Van~Engelen$^{33}$}
\author{J.~Van Lanen$^4$}
\author{E.~J.~Wollack$^{34}$}
\author{Z.~Xu$^{9,35}$}
\affiliation{$^1$ Department of Physics, Cornell University, Ithaca, NY 14853, USA}
\affiliation{$^2$ Department of Astronomy, Cornell University, Ithaca, NY 14853, USA}
\affiliation{$^{3}$ Center for Computational Astrophysics, Flatiron Institute, New York, NY, USA 10010}
\affiliation{$^4$ NIST Quantum Sensors Group, 325 Broadway, Boulder, CO 80305}
\affiliation{$^5$ Physics Department, Sapienza University of Rome, Piazzale Aldo Moro 5, I-00185, Rome, Italy}
\affiliation{$^6$ Canadian Institute for Theoretical Astrophysics, University of Toronto, Toronto, ON, M5S 3H8, Canada}
\affiliation{$^7$ School of Physics and Astronomy, Cardiff University, The Parade, Cardiff, CF24 3AA, UK}
\affiliation{$^8$ Department of Applied and Engineering Physics, Cornell
University, Ithaca, NY, USA 14853}
\affiliation{$^{9}$ Department of Physics and Astronomy, University of Pennsylvania, 209 South 33rd Street, Philadelphia, PA, USA 19104}
\affiliation{$^{10}$ Joseph Henry Laboratories of Physics, Jadwin Hall, Princeton University, Princeton, NJ, USA 08544}
\affiliation{$^{11}$ Department of Astrophysical Sciences, Peyton Hall, Princeton University, Princeton, NJ USA 08544}
\affiliation{$^{12}$ Instituto de Astrof\'isica and Centro de Astro-Ingenier\'ia, Facultad de F\'isica, Pontificia Universidad Cat\'olica de Chile, Av. Vicu\~na Mackenna 4860, 7820436, Macul, Santiago, Chile}
\affiliation{$^{13}$ Lawrence Berkeley National Laboratory, One Cyclotron Road, Berkeley, CA 94720, USA}
\affiliation{$^{14}$ Berkeley Center for Cosmological Physics, UC Berkeley, CA 94720, USA}
\affiliation{$^{15}$ Department of Physics and Astronomy, University of Pittsburgh, Pittsburgh, PA, USA 15260}
\affiliation{$^{16}$ Department of Physics, Columbia University, New York, NY, USA 10027}
\affiliation{$^{17}$ Astrophysics Research Centre, University of KwaZulu-Natal, Westville Campus, Durban 4041, South Africa}
\affiliation{$^{18}$ David A. Dunlap Department of Astronomy and Astrophysics, University of Toronto, 50 St. George St., Toronto, ON M5S 3H4, Canada}
\affiliation{$^{19}$ Dunlap Institute for Astronomy and Astrophysics, University of Toronto, 50 St. George St., Toronto, ON M5S 3H4, Canada}
\affiliation{$^{20}$ Department of Physics, Florida State University, Tallahassee FL, USA 32306}
\affiliation{$^{21}$ Department of Physics and Astronomy, Rutgers, the State University of New Jersey, 136 Frelinghuysen Road, Piscataway, NJ 08854-8019, USA}
\affiliation{$^{22}$ Department of Physics, Yale University, New Haven, CT 06511, USA}
\affiliation{$^{23}$ Centre for the Universe, Perimeter Institute for Theoretical Physics, Waterloo, ON, N2L 2Y5, Canada}
\affiliation{$^{24}$ Department of Physics, University of Chicago, Chicago, IL 60637, USA}
\affiliation{$^{25}$ Department of Astronomy and Astrophysics, University of Chicago, 5640 S. Ellis Ave., Chicago, IL 60637, USA}
\affiliation{$^{26}$ Kavli Institute for Cosmological Physics, University of Chicago, 5640 S. Ellis Ave., Chicago, IL 60637, USA}
\affiliation{$^{27}$ Enrico Fermi Institute, University of Chicago, Chicago, IL, 60637, USA}
\affiliation{$^{28}$ School of Mathematics, Statistics and Computer Science, University of KwaZulu-Natal, Westville Campus, Durban 4041, South Africa}
\affiliation{$^{29}$ Department of Physics, University of Milano-Bicocca, Piazza della Scienza 3, 20126 Milano (MI), Italy}
\affiliation{$^{30}$ Department of Physics and Astronomy, Haverford College, Haverford, PA, USA 19041}
\affiliation{$^{31}$ Department of Physics, California Institute of Technology, Pasadena, CA 91125, USA}
\affiliation{$^{32}$ Instituto de F\'isica, Pontificia Universidad Cat\'olica de Valpara\'iso, Casilla 4059, Valpara\'iso, Chile}
\affiliation{$^{33}$ School of Earth and Space Exploration, Arizona State University, Tempe, AZ, USA 85287}
\affiliation{$^{34}$ NASA Goddard Space Flight Center, Greenbelt MD 20771}
\affiliation{$^{35}$ MIT Kavli Institute, Massachusetts Institute of Technology, Cambridge, MA, USA 02139}
\label{firstpage}

\begin{abstract}
\clearpage

We present measurements of the average thermal Sunyaev Zel'dovich (tSZ) effect from optically selected galaxy groups and clusters at high signal-to-noise (up to 12$\sigma$) and estimate their baryon content within a 2.1$^\prime$ radius aperture. Sources from the Sloan Digital Sky Survey (SDSS) Baryon Oscillation Spectroscopic Survey (BOSS) DR15 catalog overlap with 3,700 sq.\ deg.\ of sky observed by the Atacama Cosmology Telescope (ACT) from 2008 to 2018 at 150 and 98 GHz (ACT DR5), and 2,089 sq.\ deg.\ of internal linear combination component-separated maps combining ACT and \textit{Planck} data (ACT DR4). The corresponding optical depths, $\bar{\tau}$, which depend on the baryon content of the halos, are estimated using results from cosmological hydrodynamic simulations assuming an AGN feedback radiative cooling model. We estimate the mean mass of the halos in multiple luminosity bins, and compare the tSZ-based $\bar{\tau}$ estimates to theoretical predictions of the baryon content for a Navarro–Frenk–White profile. We do the same for $\bar{\tau}$ estimates extracted from fits to pairwise baryon momentum measurements of the kinematic Sunyaev-Zel'dovich effect (kSZ) for the same data set obtained in a companion paper. We find that the $\bar{\tau}$ estimates from the tSZ measurements in this work and the kSZ measurements in the companion paper agree within $1\sigma$ for two out of the three disjoint luminosity bins studied, while they differ by 2-3$\sigma$ in the highest luminosity bin. The optical depth estimates account for one third to all of the theoretically predicted baryon content in the halos across luminosity bins. Potential systematic uncertainties are discussed. The tSZ and kSZ measurements provide a step towards empirical Compton-$\bar{y}$-$\bar{\tau}$ relationships to provide new tests of cluster formation and evolution models. 

\end{abstract}

\maketitle

\section{Introduction}\label{sec:intro}

The thermal Sunyaev-Zel'dovich (tSZ) effect is a consequence of cosmic microwave background (CMB) photons inverse-Compton scattering off electrons in hot, ionized gas, especially that in the intra-cluster medium (ICM) of galaxy clusters and groups, resulting in a shift in the CMB blackbody spectrum \cite{SZ69,SZ72}. An SZ effect arising from the motion of these groups and clusters, the kinematic SZ (kSZ) effect, has a different spectral signature and is an order of magnitude smaller in amplitude than the tSZ signal, making it significantly more difficult to detect in maps of the CMB. Together, the SZ effects encode rich information about galaxy groups.

The distortion in the CMB due to the tSZ effect depends on the optical depth of the gas, $\bar{\tau}$, as well as the electron temperature, $T_e$, and is proportional to the Compton-$y$ parameter. In its sensitivity to the cluster's integrated line-of-sight pressure profile, the tSZ effect is a valuable, largely redshift-independent probe of gas in the ICM. Measurements of the tSZ effect allow us to study the  thermodynamics of the cluster gas, including processes such as  active galactic nuclei (AGN) feedback, star formation, radiative cooling, and cluster merger histories. The tSZ effect can also give us information about the shapes and extents of cluster gravitational potential wells and dark matter halos \cite{2016arXiv160702442B,Mroczkowski2019}. By tracing the electron distribution within groups and clusters, the tSZ effect is sensitive to the poorly understood spatial distribution of ionized gas and the baryon content. We are able to estimate the optical depth of the gas within halos by combining tSZ measurements with cosmological hydrodynamic simulations \cite{2016arXiv160702442B}. In the case of the ``missing baryon" problem, observations suggest that this gas contains fewer baryons than would be predicted by mass-density profile models neglecting heating processes \cite{Fukugita2004,Nicastro2008,Shull2012,Rasheed3487,Chiu_2018}. These baryons are thought to be located at the outskirts of groups in the diffuse warm-hot intergalactic medium, where they are not easily measured by X-ray observations or the tSZ effect \cite{Fukugita2004,Schaan2016,Ho2009,Richter2005,Ostriker2006,ChavesMontero2019}.     

\begin{figure*}[ht!]
\begin{center}  
\hspace*{-0.75cm}
\includegraphics[width=17.2cm, trim={0 0 0 2.0cm},clip]{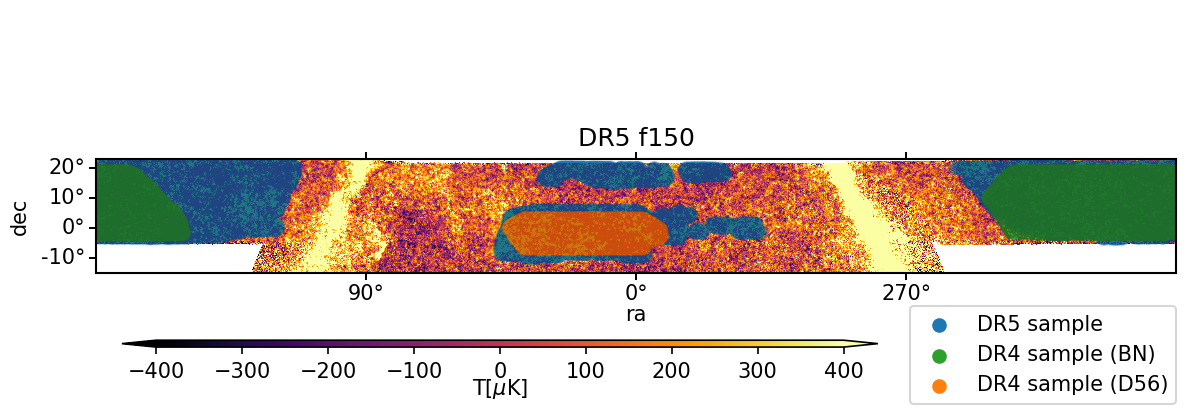}
\end{center}
\end{figure*}

\begin{figure*}[ht!]
\begin{center}
\vspace*{-0.5cm}
\hspace*{-0.75cm}
\includegraphics[width=17.2cm, trim={0 0 0 2.0cm},clip]{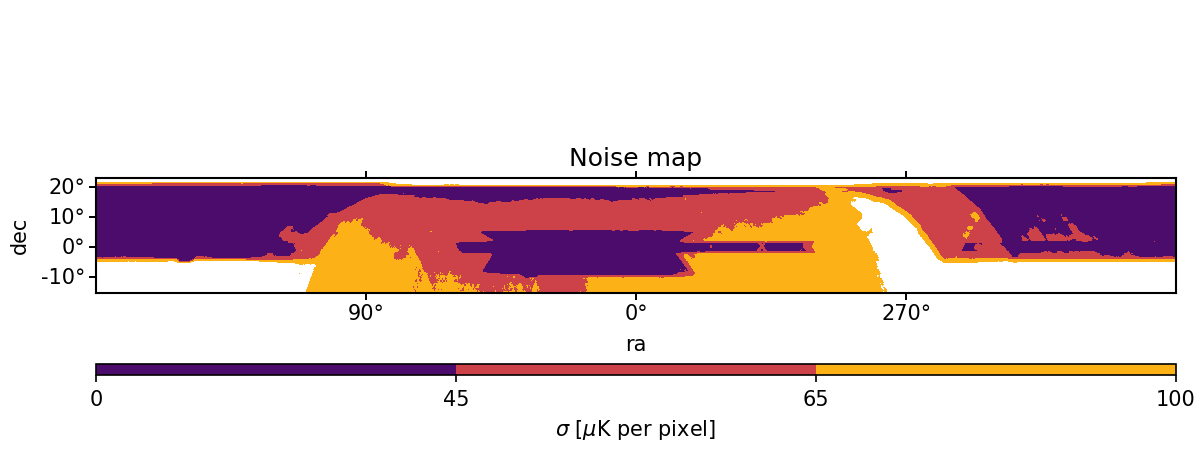}
\caption{ 
Top: The ACT~+~\textit{Planck} map used for the DR5 f150 analysis with the overlapping 343,647 SDSS DR15 selected sources plotted in blue over 3,700 sq.\ deg., and the BN and D56 areas covered by the ILC maps plotted in green and orange, respectively. Bottom: The inverse white noise variance map associated with the {\seof} coadded ACT+\textit{Planck} map highlighting regions representing a noise equivalence of 45 and 65 $\mu$K per pixel (with a 0.5 arcmin resolution plate Carr\'e projection), which were used to cut the SDSS sample for the {\seof} analysis. The orange and yellow regions of higher noise overlapped with 27$\%$ of the DR15 sample. Results are shown for the more conservative 45 $\mu$K per pixel inverse white noise variance map cut, shown in purple. We performed an equivalent cut for the {\sen} map and analysis.}
\label{fig:catalogs}
\end{center}
\end{figure*}

While the kSZ effect has an amplitude proportional to both the total cluster gas mass and the cluster's line-of-sight velocity, it is independent of the gas temperature. When both the tSZ and kSZ effects are measured for the same sample of sources and the optical depth (and thus the total cluster gas mass) is modeled from the tSZ data, the combination can be used to convert the pairwise momentum \cite{2012PhRvL.109d1101H} measured from the kSZ effect into pairwise velocity, which in turn can be used to constrain cosmological parameters, such as the sum of the neutrino masses \cite{Mueller:2014dba,Mueller:2014nsa}.

To use tSZ measurements to probe cluster properties, we must extract the tSZ signal from the microwave sky at high significance. Cleaning the tSZ signal from sources of contamination such as competing astrophysical and cosmological signals, instrumentation and atmospheric noise, and dusty galaxy and synchrotron emission can be a challenge. Recent efforts by the Atacama Cosmology Telescope (ACT) \cite{Marriage2011, Hasselfield:2013wf,Hilton2018,Aghanim2019}, South Pole Telescope (SPT) \cite{Staniszewski2009, BleemL2015}, and \textit{Planck} Collaborations \cite{planck2018results} have utilized wide-field multi-frequency data to produce measurements of the tSZ effect in CMB data. Heritage for both tSZ stacking and multifrequency dust reconstruction in ACT data includes multiple recent publications \cite{Fuzia2020,Gralla2014,Su2017}.

In this work we use multi-frequency CMB temperature maps \cite{Naess2020} (Figure \ref{fig:catalogs}) from ACT (Data Release 5, DR5) and \textit{Planck} data, along with component-separated Compton-$y$ maps \cite{Madhavacheril2019} (DR4) to extract our tSZ signals. We perform aperture photometry (AP) on stacked $18^\prime \times 18^\prime$ map cutouts (submaps) centered on sources from the Sloan Digital Sky Survey (SDSS) Baryon Oscillation Spectroscopic Survey (BOSS DR15 \cite{BOSSDR15}) in luminosity selected bins. We bin by luminosity to target the highest signal-to-noise SZ measurements and enable comparisons between independent bins of sources with differing average masses. We select 343,647 luminous red galaxies (LRGs) from the DR15 catalog based on an inverse white noise variance map cut, point source mask, and \textit{Planck} Galactic plane mask used in the production of the 2015 \textit{Planck} Compton-$y$ map \cite{Planck2015ymaps}. We then clean the tSZ signals by removing contaminating emission from dusty galaxies. We convert these tSZ signals to estimates of optical depth, $\bar{\tau}$, by using a hydrodynamical model \cite{2016arXiv160702442B}. 

\begin{figure}[!t]
\begin{center}
\hspace*{-0.25cm}
\includegraphics[width=8.6cm]{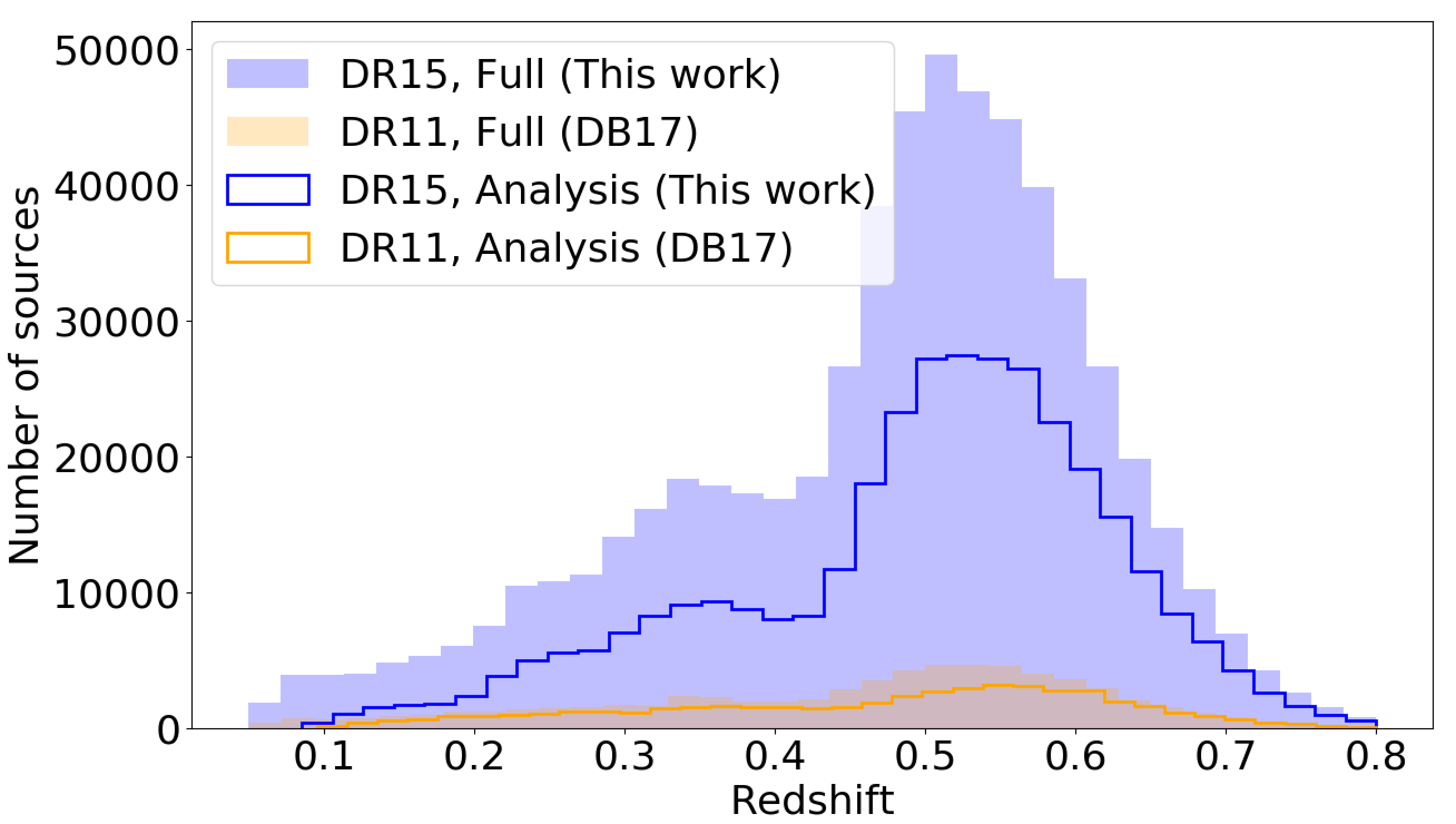}
\caption{The SDSS DR15 redshift distribution for the 602,461 total galaxy sample and selected 343,647 galaxy sample for analysis overlapping with the ACT+\textit{Planck} DR5 map, as compared to the $\sim$9 and $\sim$7 times fewer DR11 galaxies overlapping with the ACT DR3 area and those used for the 2017 result (DB17), respectively.}
\label{fig:redshift}
\end{center}
\end{figure}

In a companion paper, Calafut et al. 2020 (henceforth C21) \cite{C21}, we report a measurement of the pairwise kSZ signal using the same datasets described here to measure the kSZ effect at $>5\sigma$, a significant improvement over our previous results (De Bernardis et al.~2017, henceforth DB17 \cite{FDB2017}).  Previous measurements using the pairwise estimator \cite{Ferreira:1998id} have been reported by the \textit{Planck} collaboration using galaxies from SDSS \cite{PlanckkSZ}, and the South Pole Telescope collaboration using galaxies from the Dark Energy Survey \cite{Soergel2016}. In C21, we obtain optical depth estimates by fitting to an analytical kSZ signal model. In this work, we compare these estimates to $\bar{\tau}$ estimates obtained through tSZ measurements.

In 2011, Hand et al.~presented a measurement of the tSZ effect in ACT maps using a matched-filtering method with an assumed profile \cite{Hand2011}. In DB17, a 600 deg$^2$ map from ACT Data Release 3 (DR3) was analyzed in combination with sources from the SDSS DR11 galaxy catalog (Figure \ref{fig:redshift}) also using a matched filter for the tSZ signal extraction.  Then $\bar{\tau}$ was estimated from the same hydrodynamical simulations used in this work. In this work, we adopt the aperture photometry approach for both the tSZ and kSZ analysis and estimate both the tSZ uncertainties and kSZ covariance matrix using jackknife estimates for consistency. The tSZ results presented here are consistent with those in DB17, with smaller uncertainties. The details of the kSZ analysis that differ between this work and DB17 are discussed in C21. Although our galaxy samples are different than those in Hand et al.~2011 \cite{Hand2011}, our results are statistically consistent with those results for bins with similar average luminosity.  

A pair of contemporary papers from ACT, Amodeo et al.~2020 (A20) \cite{Amodeo20} and Schaan et al.~2020 (S20) \cite{schaan20}, use the same ACT DR5 maps \cite{Naess2020} or component-separated maps \cite{Madhavacheril2019} as are used here and in C21,  but with a different galaxy sample,  to focus on the radial dependence of the kSZ and tSZ signals, i.e., on the baryon profiles. Rather than the pairwise estimator used in C21, S20 uses the velocity reconstruction kSZ estimator, which is convenient for measuring the baryon density profile.  To obtain a complete picture of the gas thermodynamics, A20 and S20 stack on the CMASS and LOWZ galaxy samples \cite{BOSSDR10}, for which clustering and galaxy lensing measurements are available.  Because the galaxy samples are different in this work than in A20 and S20, and have different host halo masses, the results from these papers are not directly comparable to ours; however the rough signal-to-noise ratios obtained are comparable.  The promise of the pairwise kSZ estimator used in C21 lies in its potential for measurements of the pairwise velocity, if the optical depths of the sources are known.  In this paper, we investigate methods of constraining the optical depth. Overall, these various papers are complementary, and highlight the wealth of information in joint kSZ and tSZ measurements.\\
\indent In Section \ref{sec:data} we describe the ACT+\textit{Planck} maps we analyze, and discuss the selection and binning of the SDSS data used for this work. In Section \ref{sec:analysis} we summarize our aperture photometry filtering approach towards extracting tSZ signals from our maps and measuring the average Compton-$y$ within our 2.1$^{\prime}$ aperture, $\bar{y}$. We describe how we use a scaling relation from hydrodynamic simulations to convert these measurements of $\bar{y}$ into estimates of halo optical depth, $\bar{\tau}$. We explore dust and beam corrections and possible systematic effects in Section \ref{sec:sys}. We calculate theoretical estimates for these halos in Section \ref{sec:theorytau} in order to compare with our estimates. In C21, kSZ-based estimates of $\bar{\tau}$ are obtained for the same dataset. We discuss this in Section \ref{sec:kSZsignals}. We present the tSZ $\bar{y}$ measurements for each bin and estimated $\bar{\tau}$ values from the tSZ and kSZ results in Section \ref{sec:results}. The optical depth estimates from the kSZ signals from C21 and tSZ signals from this work are compared with the theoretical estimates and with one another in Section \ref{sec:taucompare}. We assume the \textit{Planck} cosmology for a flat universe \cite{Planck2015ymaps}: $\Omega_bh^2 = 0.02225$, $\Omega_ch^2 =0.1198$, $H_0 = 67.3$ km/s/Mpc, $\sigma_8 = 0.83$, $n_s = 0.964$. Combining the SZ measurements provides a step towards using observations to refine our models of galaxy formation and feedback, leading towards an empirical $\bar{y}$-$\bar{\tau}$ relationship free of the assumptions required in current simulations. 

\section{Data}\label{sec:data}

\begin{table*}
\begin{center}

\begin{tabular}{c|c|c|c|c|c|c|c|c|c|}
\cline{5-10}
 \multicolumn{4}{c|}{\multirow{1}{*}{}} & \multicolumn{3}{c|}{{\seof}, {\sen}} &
    \multicolumn{3}{c|}{\ILC}\\
\cline{5-10}
Bin & Luminosity cut/$10^{10} L_{\odot}$ & $M_{\rmn{vir}}$ cut/$10^{13} M_{\odot}$ & $\langle M_{*} \rangle/10^{11} M_{\odot}$ & N & $\langle L\rangle/10^{10} L_{\odot}$ & $\langle z\rangle$ & N & $\langle L\rangle/10^{10} L_{\odot}$ & $\langle z\rangle$ \\
\hline
L43$^*$ & $L > 4.30$  &  $M > 0.52$  & 2.21   & 343647 & 7.4 & 0.49 &   190551  & 7.4 &  0.50\\
L61$^*$ & $L > 6.10$  &  $M > 1.00$  & 2.61   & 213070 & 8.7 & 0.51 &  118852  & 8.7 &  0.51\\
L79$^*$ & $L > 7.90$  &  $M > 1.66$  & 3.17   & 103159 & 10.6 & 0.53 &  57828  & 10.6 &  0.54\\
L98     & $L > 9.80$  &  $M > 2.59$  & 3.84   & 46956  & 12.8 & 0.56 &  26308  & 12.8 &  0.57\\
L116    & $L > 11.60$ &  $M > 3.70$  & 4.50   & 23504  & 15.0 & 0.58 &   13277  & 15.0 &  0.59\\
\hline
L43D$^*$ & $4.30 < L < 6.10$  &  $0.52 < M < 1.00$  & 1.57  & 130577 & 5.2 & 0.48 &   71699   & 5.2 & 0.48 \\
L61D$^*$ & $6.10 < L < 7.90$  &  $1.00 < M < 1.66$  & 2.08   & 109911 & 6.9 & 0.48 &  61024    & 6.9 & 0.48 \\
L79D & $7.90 < L < 9.80$  &  $1.66 < M < 2.59$  & 2.61   & 56203 & 8.7 & 0.51 &  31520   & 8.7 & 0.52 \\
L98D & $9.8 < L  < 11.60$ &  $2.59 < M < 3.70$  & 3.18   & 23452 & 10.6 & 0.54 & 13031   & 10.6 & 0.55 \\

\hline
\end{tabular}
\caption{Luminosity bin labels (the $^*$bins are also analyzed in C21) and cuts (Appendix \ref{sec:lumbinning}), equivalent halo mass cuts, average stellar mass per bin, Number of sources (N), average luminosity ($\langle L\rangle$), and average redshift ($\langle z\rangle$) per luminosity bin for the final DR15 samples used in the {\seof}, {\sen}, and ILC Compton-$y$ map analyses. These samples have the noise cut, point source masks and Galactic plane mask applied. The samples differ between the coadded and ILC maps due to the difference in footprints of the two maps, with the ILC maps covering a smaller area on the sky.}\label{sourcetable}
\end{center}
\end{table*}

\subsection{ACT data}

This analysis uses a component separated internal linear combination (ILC) Compton-$y$ map, referred to as the {\ILC} map, covering the BN (1633 sq.~deg.) and D56 (456 sq.\ deg.) regions (see Figure \ref{fig:catalogs}).  The ILC map is comprised of two seasons of observations with the ACTPol receiver \cite{ThorntonACTPol} (ACT DR4) \cite{Madhavacheril2019}. In C21, the {\ILC} map used for the kSZ analysis is a CMB+kSZ map. The ILC approach was designed to account for the anisotropic noise found in ground-based CMB experiments. The ILC method combines multi-frequency data from \textit{Planck} and ACT and constructs wide-area, arcminute-resolution component-separated maps of CMB temperature and the tSZ effect. CIB-deprojected Compton-$y$ maps are also available, but were not analyzed in this work due to their higher noise. Additionally, ACT DR5 single frequency data at  150 and 98 GHz is coadded with data from \textit{Planck} at 100 and 143 GHz to cover about 21,100~sq.~deg.~of sky (3,700~sq.~deg.~of which overlaps with the SDSS data sample) \cite{Naess2020}. These maps are referred to as the {\seof} and {\sen} maps. While the DR5 data has the highest signal-to-noise ratio, the DR4 ILC results serve to check consistency with a map that already combines the multi-frequency information. A higher signal-to-noise ILC map for the DR5 data is not yet available and so was not analyzed here, but will be useful for future measurements. The beams have FWHM = 1.3$^\prime$, 2.1$^\prime$, and 1.6$^\prime$ for the {\seof}, {\sen} and {\ILC} maps, respectively, with associated uncertainties of several percent \cite{Naess2020}.

The inverse white noise variance map associated with the DR5 data is used for cutting the SDSS data sample (Section II B) and for weighting (Section III B). For the ILC analysis, the DR5 f090 inverse white noise variance map is used for weighting as it best estimates the noise properties in the ILC map. Figure \ref{fig:catalogs} shows the ACT+\textit{Planck} map overlaid with the selected 343,647 sources from the SDSS DR15 catalog. 

For the tSZ measurements, submaps are stacked on the locations of galaxies from the public Sloan Digital Sky Survey Large Scale Structure DR15 catalog\footnote{\url{https://www.sdss.org/dr15/}} from the BOSS survey \cite{2013AJ....145...10D}. The tSZ signal from the CMB map is measured at the positions of these objects. We assume that higher luminosity optical galaxies trace more massive halos. Systematic effects associated with this assumption are discussed in the context of these data in C21, and  have been previously explored for precursor data in 
\cite{Calafut2017}. The assumption that each galaxy traces one halo is discussed in Section \ref{sec:sys}.

\subsection{SDSS data}
\label{sec:sdssdata}

The 602,461 DR15 LRGs overlapping with the {\seof} ACT+\textit{Planck} map were selected for use in this analysis through a luminosity cut, a cut based on the CMB map noise level, point source masks, and a Galactic plane mask. The catalog was downloaded from the SDSS SkyServer using the query presented in Appendix \ref{sec:query}. The full catalog, with flags for the cuts described below, is publicly available.\footnote{\url{https://github.com/evavagiakis/V21_Catalog/}} The luminosities of the sources are calculated based on their (multiband) de-reddened SDSS composite model magnitudes and K-corrected using the \url{k_correct}\footnote{\url{http://kcorrect.org}} \cite{Blanton_2007} software  according to the luptitude to flux conversion outlined in \cite{Lupton_1999}. The DR15 luminosities of the selected sources range from $4.30\times10^{10}L_{\odot}$ to $2.61\times10^{12}L_{\odot}$ with an average luminosity of $7.38\times10^{10}L_{\odot}$. Luminosity bins for joint tSZ and kSZ analyses with C21 were chosen to match two of the luminosity cuts from DB17 ($L=7.9\times10^{10} L_{\odot}$ and $L=6.1\times10^{10} L_{\odot}$) as well as one lower luminosity cut ($L=4.3\times10^{10} L_{\odot}$). The three disjoint bins based on these cuts were selected to have roughly equal spacing, and such that each bin has over 100,000 galaxies that pass cuts for analysis with the DR5 maps (Table~\ref{sourcetable}). The cumulative luminosity bins include the highest signal-to-noise SZ measurements, while the disjoint bins enable comparisons between independent bins.  Since the tSZ signal-to-noise ratio is higher for high mass halos, we also perform the tSZ analysis for the two highest mass bins from DB17 (L98 and L116) which are not studied in C21. A plot of the luminosity distribution and bins is shown in Appendix~\ref{sec:lumbinning}. 

After we selected luminosity bins and applied the minimum luminosity cut, which removed 80,162 lower luminosity sources from the sample, we performed an inverse white noise variance cut. Figure \ref{fig:catalogs} shows the inverse white noise variance map that was used to explore cuts in the DR15 sample. Initially two different cuts were studied (45 and 65\,$\mu$K per pixel).  The more conservative cut of 45\,$\mu$K per pixel was selected for all subsequent analyses based on a signal-blind uncertainty analysis which compared the jackknife error bars on the aperture photometry (AP) analysis for the samples cut by 45\,$\mu$K per pixel, 65\,$\mu$K per pixel, and no noise cut. The 45\,$\mu$K per pixel cut removes 27$\%$, or 140,209 sources from the overlapping sample and improves the jackknife error bars as compared to no noise cut.

Galactic plane masking was then performed with the mask used in the production of the 2015 \textit{Planck} Compton-$y$ map to reduce Galactic contamination \cite{Planck2015ymaps}. The 50$\%$ mask was selected to conservatively cut sources from the Galactic plane region of the ACT+\textit{Planck} map, resulting in a cut of 26,521 additional sources from the DR15 sample. 

To mask point sources we first used the two source masks developed for Choi et al.~2020 \cite{Choi:2020}. For the lowest noise D56 region a 15\,mJy point source mask was applied, and for the higher noise regions a 100\,mJy mask was applied. Together these masks removed an additional 11,922 sources from our sample, leaving us with a final selected sample of 343,647 sources after all masks and cuts are applied.

Figure \ref{fig:redshift} shows the redshift distributions of the 343,647 DR15 galaxies overlapping with the ACT+\textit{Planck} map after the luminosity, inverse white noise variance map, Galactic plane mask and point source mask cuts. The 67,938 DR11 sources overlapping with the coadded ACT DR3 map used in our 2017 result are also shown for comparison (DB17 \cite{FDB2017}). The redshifts range from $0.08$ to $0.8$ with an average reshift of $0.49$ for the DR15 catalog.    

Histograms of the luminosities depicting cuts and bins are provided in Appendix \ref{sec:lumbinning}. The properties of each luminosity-selected bin are summarized in Table \ref{sourcetable}.  The mean stellar masses are estimated from the mean luminosities assuming $M_*/L=3.0$ as predicted by the Chabrier IMF \cite{Chabrier} and discussed in \cite{Kravtsov2014,Bell2001,Bell2003,Bernardi2010}. The mean halo masses are derived from the $M_*-M_{vir}$ relation from abundance matching as described in \cite{Kravtsov2014}. Abundance matching is a statistical technique to model the correlations between galaxy and halo properties based on mapping galaxies to dark matter halos of the same number density in the universe. 

The same mass cuts and average redshifts of the sources reported in Table \ref{sourcetable} are used when calculating the linear model prediction from the kSZ measurements in C21.

\section{Analysis}\label{sec:analysis}

\begin{figure*}[htbp]
\begin{center}
\includegraphics[width=12.9cm]{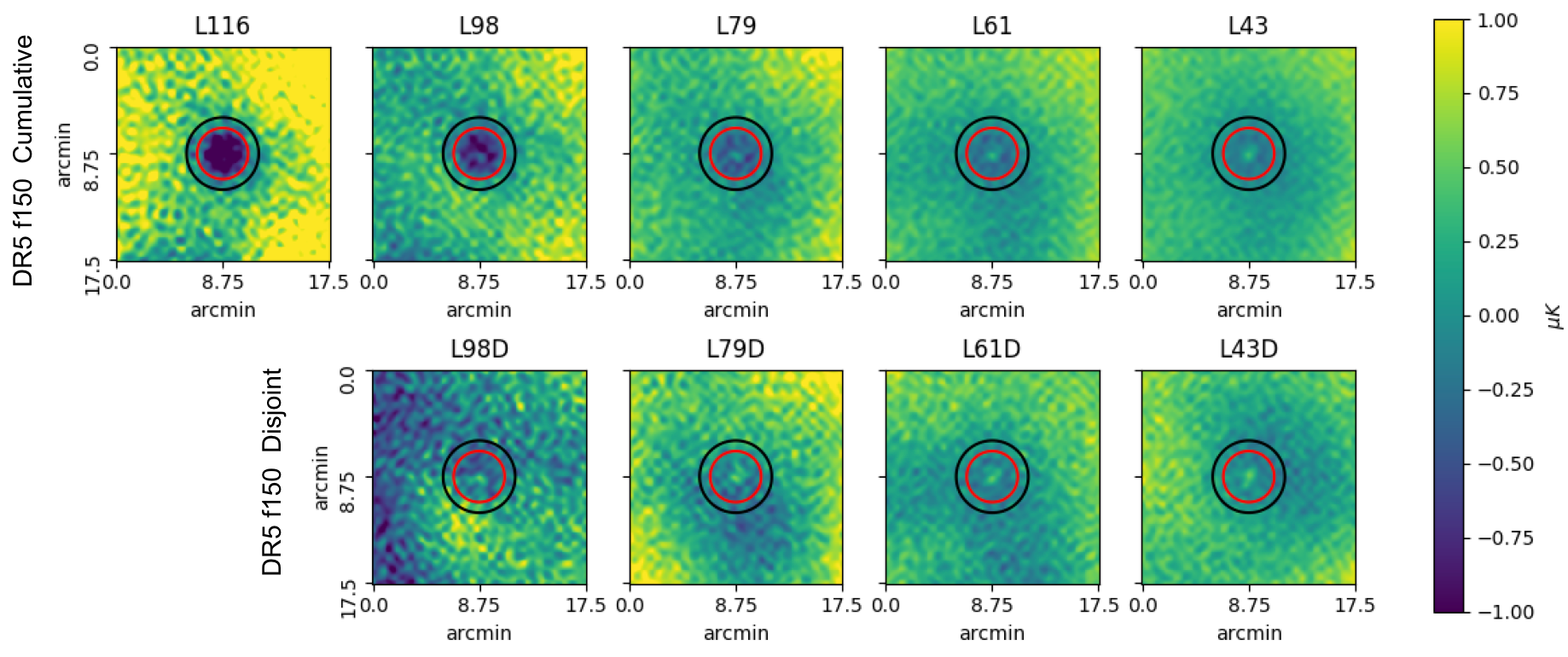}
\includegraphics[width=12.9cm]{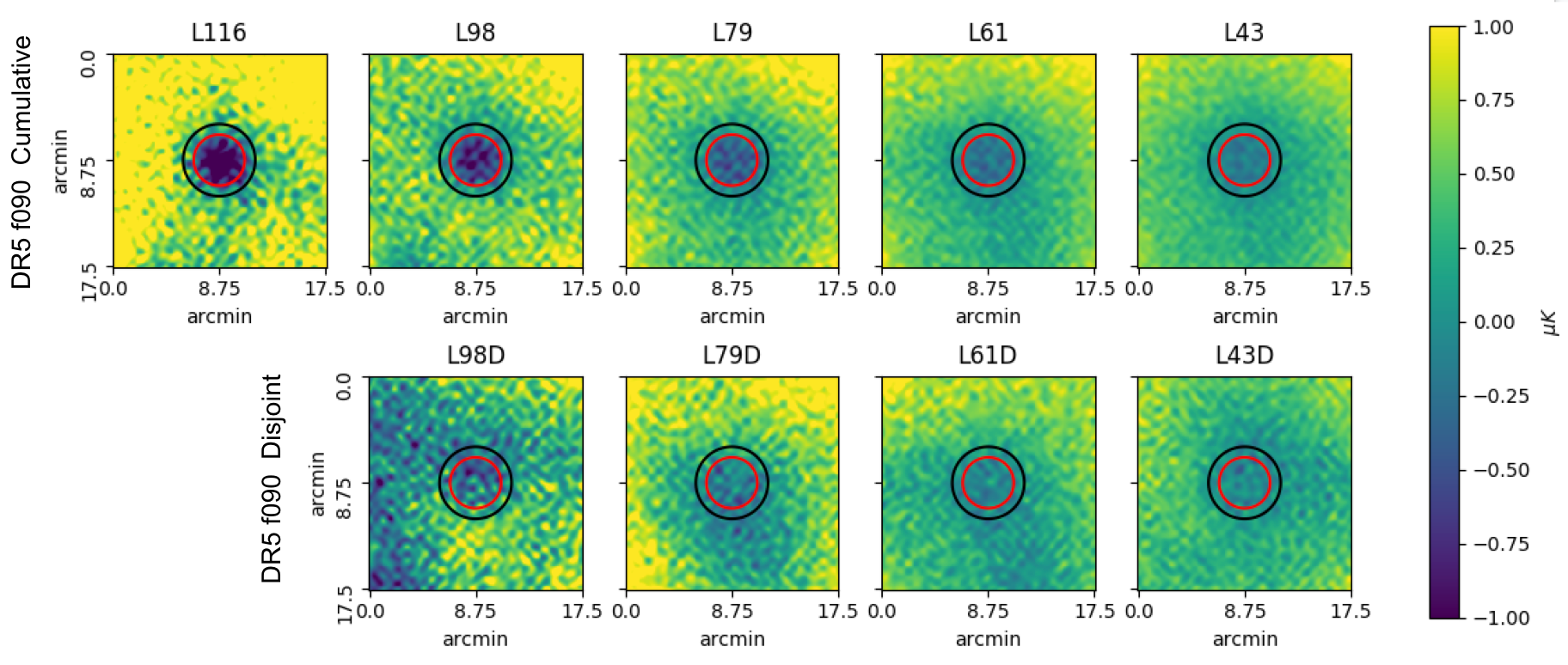}
\includegraphics[width=12.9cm]{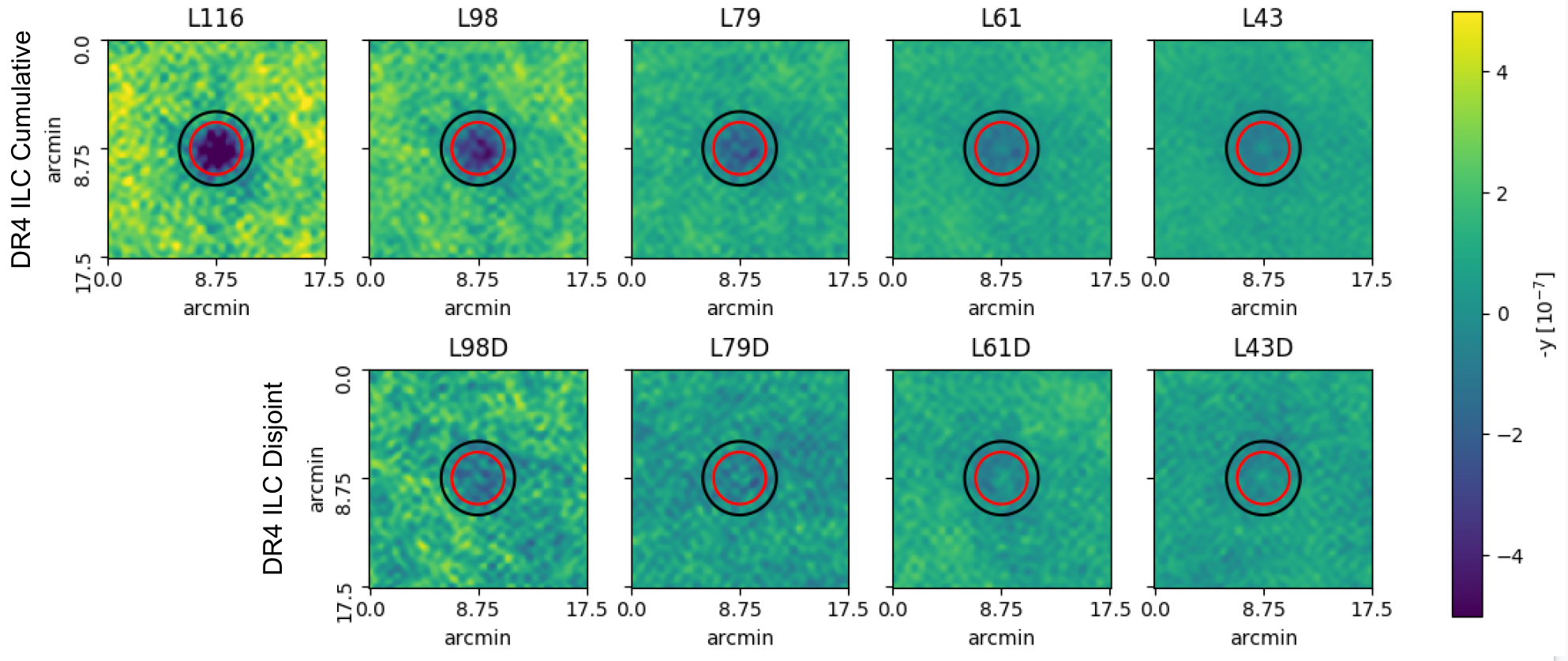}
\caption{Stacked raw submaps for the five cumulative (top) and four disjoint (bottom) luminosity bins as defined in Table \ref{sourcetable}, for the {\seof} map (top two rows, in differential units of CMB temperature which are calibrated to \textit{Planck} \cite{Naess2020}), the {\sen} map (middle two rows, in the same units as the top two rows) and the {\ILC} Compton-$y$ maps (bottom two rows, in negative units of $y$ to better compare to the coadded maps). The submaps represent the weighted average submaps of the sources in a given bin, where the weight for each source is taken to be the average value inside the accompanying $R_{1} = 2.1$ submap  in the inverse white noise variance map. The sub-0.5$^\prime$-scale structure in the submaps is an artifact of the sub-pixel interpolation and is not physical. The maps are normalized with the average value within the AP annulus, such that the mean of the pixels in the annulus in these maps is equal to zero. The apertures used for the tSZ and kSZ AP are drawn, where $R_{1} = 2.1^\prime$ (red) and $\sqrt2R_{1}$ (black). Radial averages of these submaps are plotted in Figure \ref{fig:radialavg}. A central bright spot due to dust on approximately the beam scale can be seen across luminosity bins in the DR5 f150 submaps, but not the DR5 f090 submaps. Dust contamination of the DR4 ILC maps is more subtle, but can be noticed in radial average plots (Figure \ref{fig:radialavg}).}
\label{fig:submaps}
\end{center}
\end{figure*}

\begin{figure*}[htbp]
\begin{center}
\includegraphics[width=12.9cm]{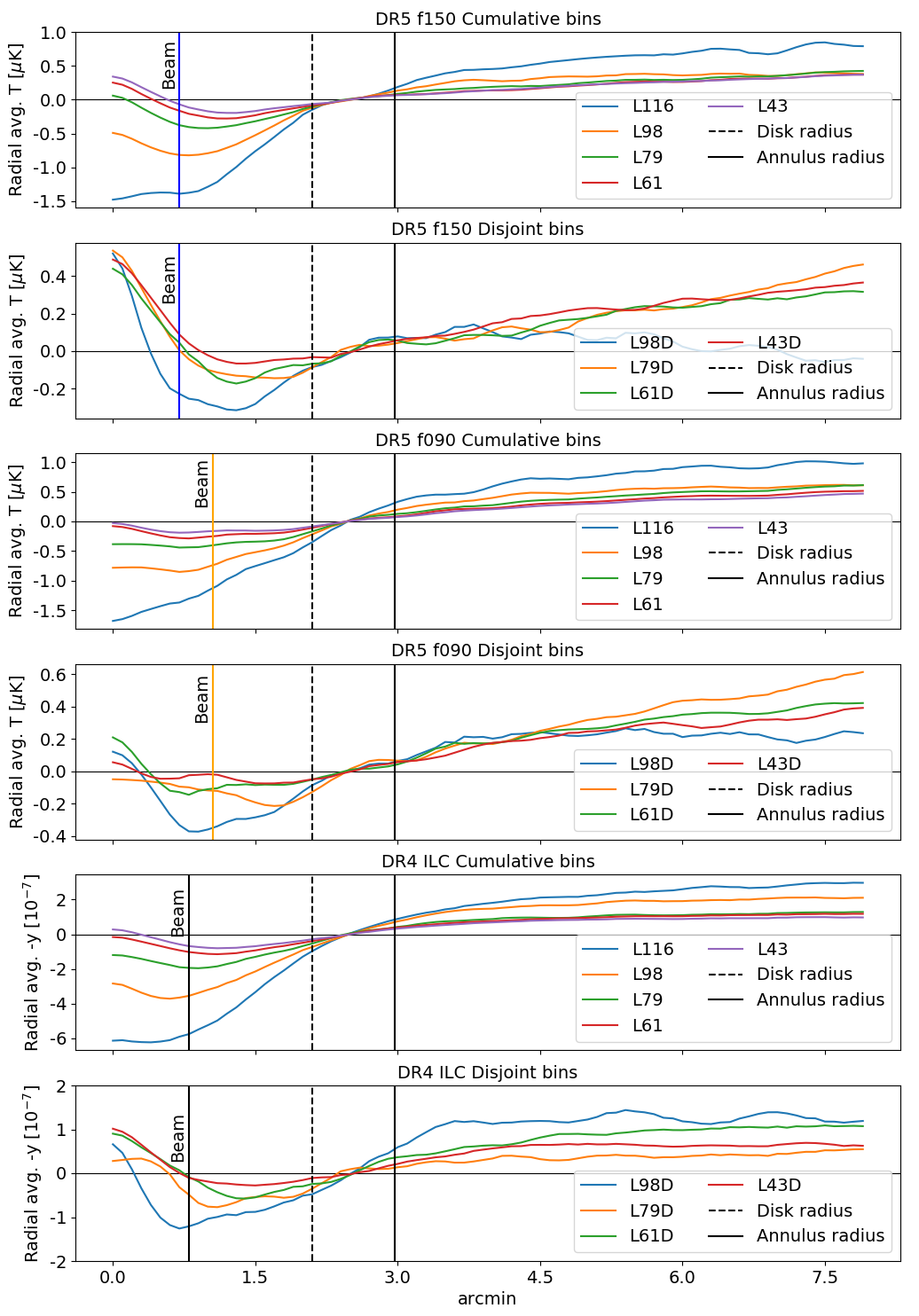}
\caption{Radial average of the stacked submaps, which have been repixelized to 0.1$^\prime$ per pixel, normalized to the average annulus value, for each luminosity bin, for the {\seof} and {\sen} coadded maps as well as the {\ILC} map, shown for illustrative purposes only. The native units of the {\seof} and {\sen} maps are in $\mu \mathrm{K}$ and the {\ILC} maps in $y$. Negative $y$ is plotted here to compare with the decrements present in temperature. The aperture photometry disk radius is plotted as a vertical black dashed line, and the annulus outer radius is plotted as a vertical solid black line. The beam radius is plotted as a blue vertical line for {\seof}, an orange vertical line for {\sen}, and a black vertical line for the effective {\ILC} beam. A central ``bright spot" is observed in nearly all but the most luminous bin, and is attributed to dust emission. Due to this effect, we studied the core-excised AP approach for the {\seof} and {\sen} analysis (Sections \ref{sec:tSZextraction}, \ref{sec:sys}).}
\label{fig:radialavg}
\end{center}
\end{figure*}

We use aperture photometry (AP) to filter the Compton-$y$ and {\seof} maps and extract the tSZ signals by stacking on source-centered submaps. The same 2.1$^\prime$ AP filter is used in the kSZ analysis in C21. This approach to comparing estimated optical depths is an improvement over the 2017 result in which a matched filter was used for the tSZ analysis while AP was used for the kSZ analysis.

\subsection{Filtering CMB maps}\label{sec:filt}

We select submaps of 18$^\prime \times 18^\prime$ (about three times larger than the outer diameter of the AP annulus) of pixel size 0.5$^\prime$, interpolate in the Fourier domain with a pixel size of 0.1$^\prime$ per pixel, and reproject to a coordinate system centered at the galaxy center position using \url{pixell}\footnote{\url{https://github.com/simonsobs/pixell}}. A comparison between different pixelization approaches is discussed in C21 Appendix A, and we use here the same approach adopted in C21. On each source-centered submap, we draw an aperture of $R_{1} = 2.1^\prime$ located at the central coordinate in RA and DEC provided by the DR15 catalog. The signal associated with the sources is taken to be the average within this aperture, minus the average of the pixels within an annulus of inner radius $R_{1} = 2.1^\prime$ and an outer radius of $\sqrt2R_{1}$. The 2.1$^\prime$ size of this annulus was selected to correspond to $\sim$0.8~Mpc based on the average angular diameter distance for the source sample and assumed cosmology (Section \ref{sec:intro}). An associated weight for each source is assigned by taking the average within a $R_{1} = 2.1^\prime$ disk centered on the same source position on a  18$^\prime \times 18^\prime$ submap taken from the inverse white noise variance map. This value is used for weighting the signals in the tSZ analysis.

\subsection{tSZ signal extraction}\label{sec:tSZextraction}

\begin{figure*}[ht!]
\begin{center}
\hspace*{-0.2cm}
\includegraphics[width=17.2cm]{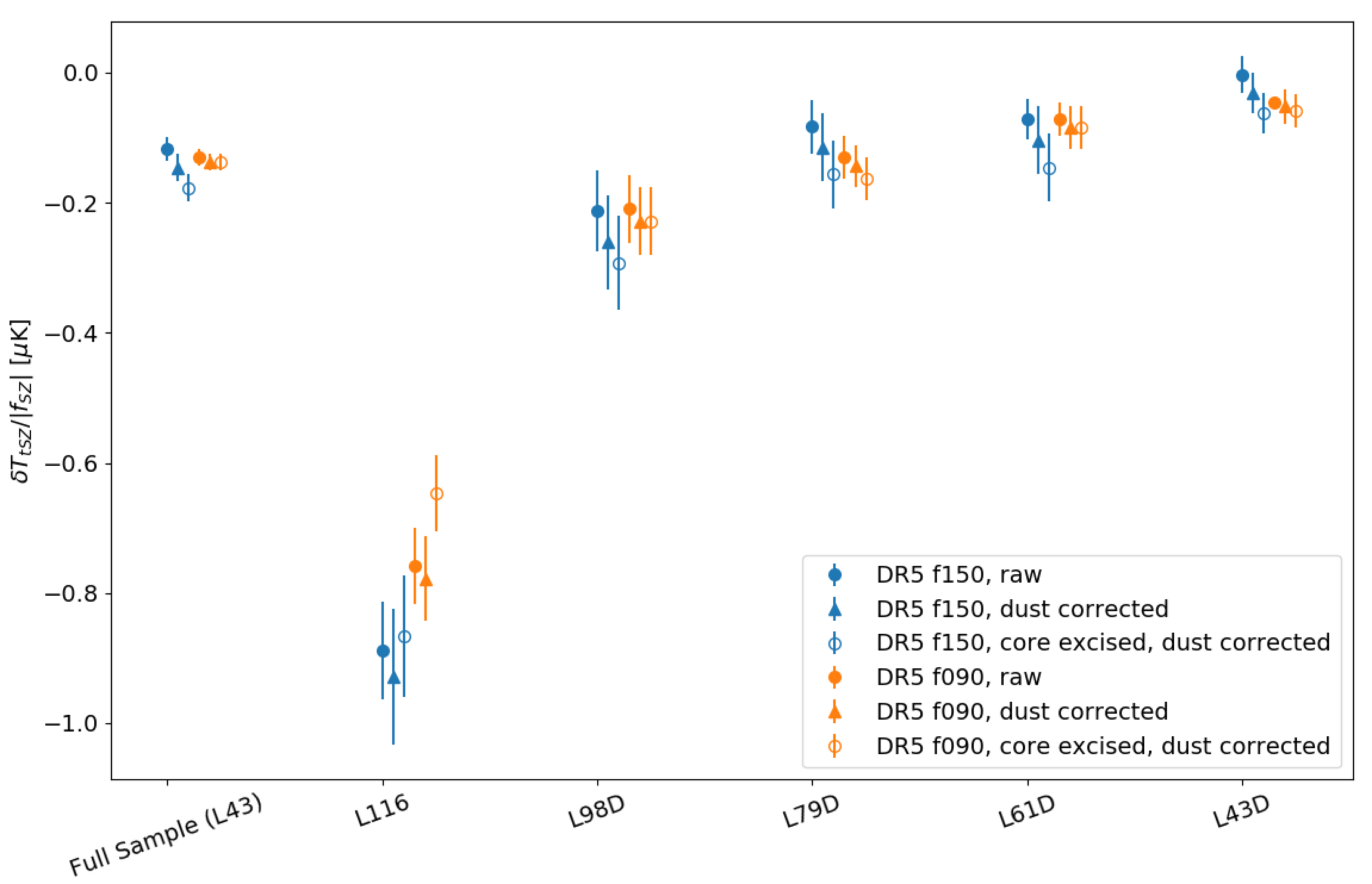}
\caption{Aperture photometry tSZ signals in units of temperature rescaled by $|f_{\rm SZ}|$ (Equation \ref{eq:fsz}) for disjoint luminosity bins and the full analysis sample, with 1$\sigma$ jackknife estimated uncertainties, for the {\seof} and {\sen} maps. Raw results using the 2.1$^{\prime}$ radius are compared to results after \textit{Herschel} dust correction (Section \ref{sec:sys}) and the removal of the pixels within a beam-scale radius (0.7$^\prime$ for the {\seof} map and 1.1$^\prime$ for the {\sen} map, Section \ref{sec:tSZextraction}). The {\sen} results shown have been beam corrected (Section \ref{sec:tSZextraction}). The \textit{Herschel} dust correction has a small effect on the temperature signals and lowers signal-to-noise for the {\sen} map due to the relatively large error bars on the dust estimates which are propagated into the uncertainties on tSZ signal. The ``core-excised" AP approach removes a noticeable amount of tSZ signal in the L116 bin, as there is less dust contamination apparent on the beam scale in the stacked submaps for this bin as compared to the lower luminosity bins (Figure \ref{fig:radialavg}). For the most part, it has less than a $1\sigma$ effect on the tSZ signals, and is not included in the rest of this analysis. The dust corrected (filled triangle) results are propagated through to our comparisons of optical depth estimates.}
\label{fig:dustcorrplot}
\end{center}
\end{figure*}

We stack on the positions of DR15 sources in the nine luminosity bins listed in Table \ref{sourcetable} to measure the average $\delta T_{SZ}$ in differential CMB temperature units from the DR5 f090 and DR5 f150 maps, and the average Compton-$y$ for the DR4 ILC map. Round number luminosity cuts were chosen to yield five cumulative and disjoint luminosity bins, before any cuts were made to the DR15 sample, or any analyses were run. The weighted averages of the stacked submaps for each bin are shown in Figure \ref{fig:submaps}, and the radial averages of these submaps are plotted in Figure \ref{fig:radialavg}. The signal associated with each source in a bin is taken to be the average of the $\langle {\rm disk} \rangle-\langle {\rm ring} \rangle$ values per source, weighted by the associated average inverse variance weight per source. The presence of dust emission on angular scales comparable to or less than the beam size is visible in all the maps analyzed, but is most prominent in the {\seof} map (Figures \ref{fig:submaps}-\ref{fig:radialavg}). We study the impact of discarding the pixels within the beam radius from our analysis, as discussed further in Section \ref{sec:sys}. This ``core-excised" AP method removes the SZ signal within the beam radius along with the dust emission; it has a small (generally $<1\sigma$) effect on the tSZ signals (Figure \ref{fig:dustcorrplot}), so it is not adopted for the final analysis. The 2.1$^{\prime}$ AP signals are then averaged in each bin to obtain a stacked tSZ signal, $\delta T_{\rm tSZ}$. The uncertainty associated with the stacked tSZ signal for each bin is obtained using a jackknife estimation method over the sources with 2,000 iterations per bin. 

We correct the $\delta T_{\rm tSZ}$ and $\bar{y}$ estimates obtained from the DR5 f090 map to account for the larger DR5 f090 beam (FWHM=2.1$^\prime$) compared to the DR5 f150 map (FWHM=1.3$^\prime$) \cite{Naess2020}. We also correct the $\bar{y}$ estimates from the {\ILC} map which has an effective beam corresponding to a 1.6$^\prime$ FWHM Gaussian \cite{Madhavacheril2019}. We compute the beam correction factors as follows: we consider a fiducial pressure profile for the average virial mass ($M_{\rmn vir}$) in each bin \cite{Battaglia+12b}; we derive three estimates of the Compton-$y$ signal in a 2.1$^\prime$ AP filter, convolved with the f150 and f090 beams from \cite{Naess2020} and with a 1.6$^\prime$ FWHM Gaussian beam, using \texttt{Mop-c GT} \footnote{\url{https://github.com/samodeo/Mop-c-GT}} (see \cite{Amodeo20} for a detailed description on how the projection from pressure to Compton-$y$ profile and the beam convolution are implemented). We find an AP beam correction of 31\% for the DR5 f090 measurements, and a beam correction of -5\% for the DR4 ILC measurements. Thus, the DR5 f090 $\delta T_{\rm tSZ}$ measurements are multiplied by 1.3, which propagates into the DR5 f090 Compton-$y$ estimates and resulting analyses, and the {\ILC} $\bar{y}$ measurements are multiplied by 0.95. The value for the correction changes negligibly across luminosity bins, and is insensitive to our estimate of $M_{\rmn vir}$ (the change in factor is $\sim1\%$ if we vary $M_{\rmn vir}$ in our mass range). While Figures \ref{fig:submaps} and \ref{fig:radialavg} show the raw data, the DR5 f090 and {\ILC} data in the rest of this work are multiplied by these factors. We note that because the f150 beam is not exactly Gaussian, but features secondary lobes, we get a negative ($<1$) ILC correction, contrary to what we would expect given the larger ILC FWHM.

For the DR5 f150 and DR5 f090 analysis, we then estimate the aperture-averaged Compton-$y$ parameter $\bar{y}$ from this temperature signal and use a theoretical relation between optical depth and $\bar{y}$ from hydrodynamical simulations to infer the optical depth \cite{2016arXiv160702442B}. To obtain the Compton-$y$ parameter we follow the steps detailed in \cite{Hasselfield:2013wf} and DB17. The tSZ temperature signal is related to the Compton-$y$ parameter by
\begin{equation}\label{y0}
\frac{\delta T_{\rm tSZ}(\theta)}{T_{\rm CMB}} = f_{\rm SZ}\hspace{3pt}y(\theta),
\end{equation}
where $y(\theta)$ is the Compton parameter at a projected angle $\theta$ from the cluster center and, in the non-relativistic limit, $f_{\rm SZ}$ depends on observed radiation frequency:
\begin{equation}\label{eq:fsz}
f_{\rm SZ} = \Big(x \frac{e^x+1}{e^x-1}-4\Big).
\end{equation}
Here, $x=h\nu/k_BT_{\rm CMB}$ \cite{Itoh2000}. The effective band centers vary based on sky position in the coadded maps, so the median values of the SZ-weighted band centers are chosen for our analysis. The appropriately weighted band center frequencies are 97.8 and 149.6 GHz for the {\sen} and {\seof} maps, respectively, so $f_{\rm SZ, f090}=-1.53$ and $f_{\rm SZ, f150}=-0.958$ \cite{H20}. Relativistic corrections are negligible for this sample mass and can safely be excluded \cite{Nozawa2006}. We assume the ACT-based band center frequencies because the ACT measurements dominate over those from $Planck$ at our scales \cite{Naess2020}. These frequencies are associated with a 2.4 GHz uncertainty, and detailed bandpass considerations could carry a larger impact (at the few percent level). However, each of these has a small effect on $\bar{y}$ compared to our uncertainties. For each luminosity bin, $\bar{y}$ is obtained from the AP temperature signal $\delta T_{\rm tSZ}$ via equation (\ref{y0}). 

A jackknife approach is used to estimate the uncertainties associated with the binned tSZ signals and the covariance matrix for the kSZ analysis. We split the sample of sources into $N$ smaller subsamples and remove one subsample at a time, calculating the weighted average for each bin on the summed $N-1$ subsamples to obtain $N$ realizations of the computation. The jackknife method has the advantage of being self-contained and not requiring external information such as simulated CMB maps. We have varied $N$ to check the convergence of the jackknife algorithm, selecting the value that provided a stable significance against variations of $N$. For the tSZ analysis, a choice of $N=2000$ was conservative.

In simulations with AGN feedback, Battaglia \cite{2016arXiv160702442B} finds the relationship between $\bar{y}$ and optical depth to be 
\begin{equation}
\label{eqn:ysim}
\ln(\bar{\tau}) = \ln(\tau_0) + m\ln(\bar{y}/10^{-5})
\end{equation}
where $\ln(\tau_0) =-6.40$ and $m$ = $0.49$ at z = 0.5. We use this to estimate $\bar{\tau}$ from our $\bar{y}$ measurements. The systematic error bars on $\bar{\tau}$ from the $\bar{y}$-$\bar{\tau}$ relationship are calculated by using the Monte Carlo method taking into account the estimated $4\%$ systematic uncertainty on $\ln(\tau_0)$ and 8$\%$ on $m$. The systematic uncertainties on $\ln(\tau_0)$ and $m$ were estimated in \cite{2016arXiv160702442B} by taking the largest relative differences between the radiative cooling and AGN feedback models used for Equation \ref{eqn:ysim}.

A null test was performed by stacking the ILC sample on a simulated ILC CMB-only map. This test served as a check of our pipeline and jacknife estimates. The same pipeline used in the ILC analysis was used to measure the Compton-$y$ parameter in the simulated CMB map for all luminosity bins in Table \ref{sourcetable}, print stacked submaps as in Figure \ref{fig:submaps}, and plot radial averages as in Figure \ref{fig:radialavg}. The Compton-$y$ results for the simulated CMB map were consistent with zero within $1\sigma$ for all luminosity bins. The stacked submaps were also consistent with the same noise level as estimated using the jackknife uncertainty estimations.

\subsection{tSZ systematic effects}\label{sec:sys}

Several systematic effects have the potential to impact the amplitude of the measured tSZ signals.
One potential systematic for the tSZ measurement is the light emitted from star-forming SDSS galaxies in the optical/UV that is absorbed by dust grains and re-emitted in the infrared/sub-mm. To account for this, for both coadded maps we fit a model of the spectral energy distribution (SED) to \textit{Herschel} data from one large extragalactic survey that overlaps with SDSS, the \textit{Herschel} Astrophysical TeraHertz Large Area Survey (H-ATLAS) \cite{eales+10}, in the three fields GAMA-9, GAMA-12 and GAMA-15. We use the three SPIRE photometric bands centered at 250 $\mu$m, 350 $\mu$m, and 500 $\mu$m. Table \ref{dusttable} reports the number of galaxies in the SDSS sample overlapping with the \textit{Herschel} map areas for each luminosity bin analyzed.

\begin{figure*}[ht!]
\begin{center}
\includegraphics[width=17.2cm]{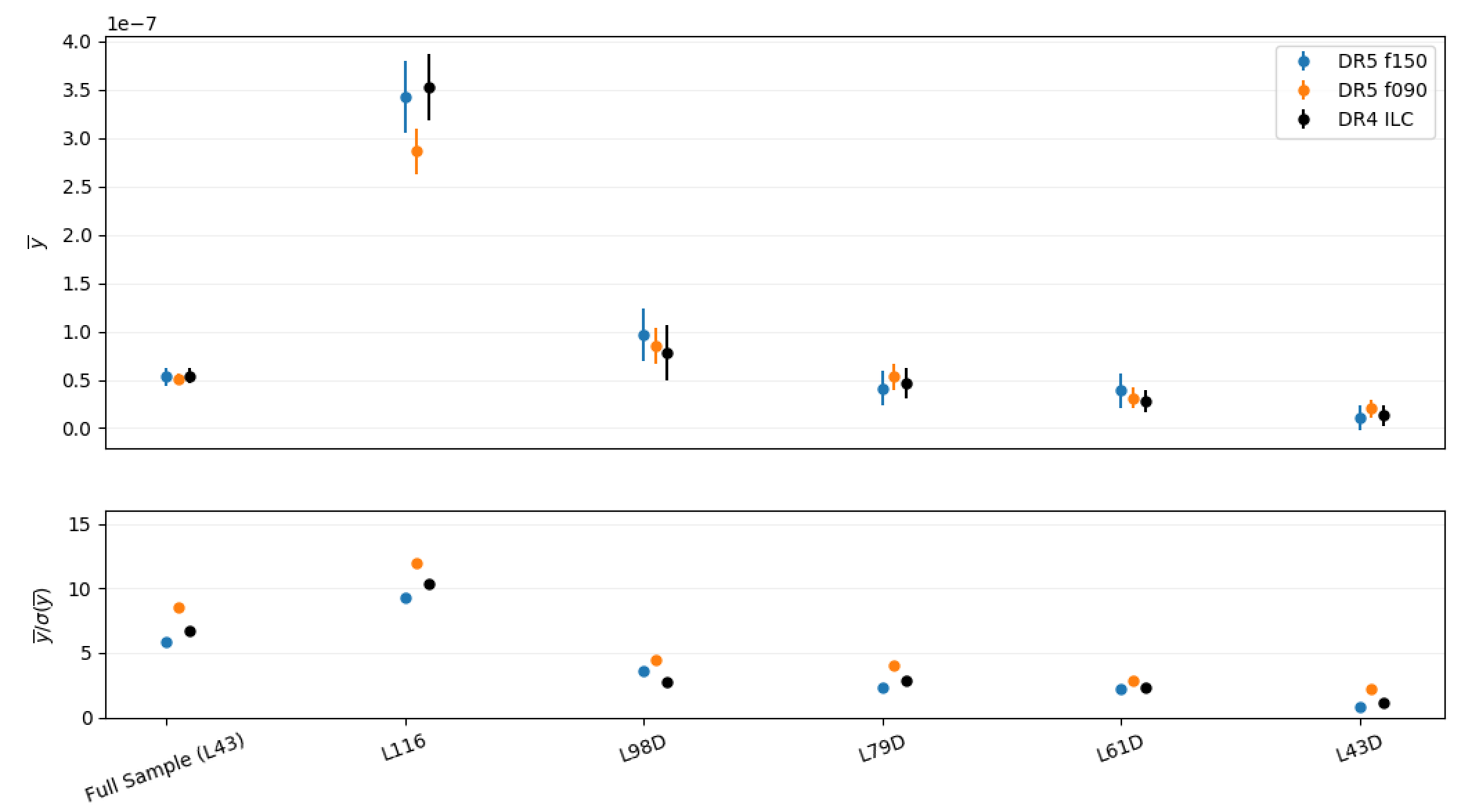}
\caption{Average Compton-$y$ in the 2.1$^\prime$ aperture for the five disjoint (highest luminosity bin L116 and bin labels with suffix ``D") luminosity bins and the full source sample ($L>4.3\times10^{10} L_{\odot}$), with jackknife estimated uncertainties, for the {\seof} and {\sen} maps after dust and beam correction (Sections \ref{sec:tSZextraction}, \ref{sec:sys}, Table 
\ref{dusttable}) and the {\ILC} Compton-$y$ map after beam correction. The lower panel shows the signals divided by their associated uncertainties. The significance of the tSZ effect observed generally decreases for less luminous sources, as expected. The results from each of the three maps are consistent.}
\label{fig:dtplots}
\end{center}
\end{figure*}

We apply an aperture photometry filter at the position of each galaxy in our SDSS sample with aperture 2.1$^\prime$ and stack the signal measured from the \textit{Herschel} maps following the approach described in A20 \cite{Amodeo20} (see their Appendix B)\footnote{While A20 measure the dust profile within apertures of increasing radii, here we measure the dust signal in a single aperture of radius 2.1$^\prime$.}. 
Using these measurements, we fit a model of the dust SED described by the following modified blackbody:
\begin{equation}
\label{eq:dustmodel}
I(\nu) =  ~A_{\rm d} \left( \frac{\nu (1+z)}{\nu_0} \right)^{\beta_{\rm d}+3} \frac{e^{(h\nu_0 / k_B T_{\rm d})}-1}{e^{(h\nu (1+z) / k_B T_{\rm d})}-1} \,,
\end{equation}
where $\nu_0 = 857$ GHz is the rest-frame frequency at which we normalize the dust emission, $A_{\rm d}$ is the amplitude of the dust emission in [kJy/sr], $\beta_{\rm d}$ is the dust spectral index, and $T_{\rm d}$ is the dust temperature in K. 
We assume flat priors for the the dust amplitude and temperature parameters in the ranges: $0.05<A_{\rm d}~{\rm [kJy/sr]}<5$, $10<T_{\rm d} ~{\rm [K]}<40$. Given the degeneracy among the parameters, we assume a Gaussian prior for the emissivity index centered on 1.2 and with standard deviation of 0.1, truncated in the range $1<\beta_{\rm d}<2.5$, as in \cite{Amodeo20}.
We obtain constraints on the model parameters using an MCMC sampler (\texttt{emcee} \cite{emcee}) to estimate the posterior probability function. Our best-fit values for the dust temperature are in the range $21<T_{\rm d} ~{\rm [K]}<30$ across the luminosity bins, with $1\sigma$ uncertainties of $\sim30$\%, and $\beta_{\rm d}=1.2\pm0.1$ consistent with our prior.  
We finally infer the amount of dust emission at 150 GHz and 98 GHz. We report our estimates of the dust emission in $\mu$K units; these need to be removed from the tSZ signal in Table \ref{dusttable}. We obtain these values by multiplying our best fit estimates from eq. \ref{eq:dustmodel},  with $\nu=150$ GHz and $\nu=98$ GHz, respectively, by the factor $\left( \frac{dB(\nu,T)}{dT} \right)^{-1}$, where $T=T_{\rm CMB}$ and $B$ is the Planck function, in order to convert from the \textit{Herschel} map intensity units to differential CMB temperature units that match the units of our SZ measurements. Because the error bars on the dust estimates are asymmetric, the largest values for each data point are selected for the more conservative estimate of $\bar{y}/\sigma(\bar{y})$ (Figure \ref{fig:dtplots}).

\begin{table}[ht!]
\setlength\extrarowheight{2pt}
\begin{tabular}{ c |c |c |c}

Bin & N & T$_{\mathrm{dust, 150 GHz}}$ [$\mu$K] & T$_{\mathrm{dust, 98 GHz}}$ [$\mu$K]\\  
\hline
L43 & 12726  & $0.026^{+0.014}_{-0.008}$ & $0.012^{+0.008}_{-0.004}$\\ 
L61 & 7784  & $0.028^{+0.019}_{-0.010}$ & $0.013^{+0.010}_{-0.005}$\\ 
L79 & 3858  & $0.029^{+0.016}_{-0.010}$ & $0.014^{+0.009}_{-0.005}$\\ 
L98 & 1795  & $0.037^{+0.027}_{-0.014}$ & $0.017^{+0.014}_{-0.007}$\\
L116 & 941  & $0.043^{+0.063}_{-0.023}$ & $0.021^{+0.031}_{-0.011}$\\  
\hline
L43D  & 4872  & $0.026^{+0.022}_{-0.011}$& $0.013^{+0.011}_{-0.005}$\\  
L61D  & 3926  & $0.034^{+0.037}_{-0.017}$& $0.018^{+0.023}_{-0.010}$\\
L79D & 2063  & $0.027^{+0.026}_{-0.012}$& $0.014^{+0.015}_{-0.007}$\\ 
L98D & 854   & $0.049^{+0.035}_{-0.021}$& $0.022^{+0.016}_{-0.010}$\\
\hline
\end{tabular}
\caption{Estimated dust signal and 1$\sigma$ statistical
uncertainties for the {\seof} and {\sen} maps using 4 percent of the SDSS sample used in this analysis overlapping with the \textit{Herschel} map areas, as computed using the method described in \cite{Amodeo20}.}\label{dusttable}
\end{table}

In the raw data, we observe a central ``bright spot" in several luminosity bins in all maps, most notably the {\seof} map. We attribute this to additional dust contamination at approximately the beam scale in the stacked submaps (Figure \ref{fig:submaps}), as shown in the radial averages presented in Figure \ref{fig:radialavg}. 

This apparent dust contamination is particularly present in the {\seof} map and is stronger than the emission estimated from \textit{Herschel}.  The apparent angular scale, comparable to the beam size, suggests a compact source.  We explored the effect of this excess emission by using a core-excised AP approach for the DR5 maps, using an aperture photometry filter which excluded  the central disk on the beam scale (Section \ref{sec:tSZextraction}). We do not include the core-excised AP approach in our reported results or optical depth comparisons due to the small (generally $<1 \sigma$) impact on the tSZ signals, to avoid biasing our results by removing the central portion of the tSZ signal, and to best compare with the results from C21. Thus, while the \textit{Herschel} dust correction accounts for some of the dust present in the {\seof} and {\sen} maps, this correction is imperfect, and some contamination remains from source galaxy emission. The effects of the dust correction steps taken on the DR5 f150 and DR5 f090 analyses can be seen in Figure \ref{fig:dustcorrplot}.

Another potential cause of increased uncertainty in the tSZ measurements is dust and synchrotron emission from our own galaxy. To account for this, we apply the same 50$\%$ Galactic plane mask as used in the production of the 2015 $Planck$ all-sky Compton-$y$ maps \cite{Planck2015ymaps}. This cut eliminated 16,977 sources after all other cuts are considered, or 6$\%$ of our final sample, and was found to have a negligible impact on our signal compared to not applying the mask. We do not expect that synchrotron emission from the SDSS galaxies themselves has a significant impact on our discussion in this work based on the core excised comparison shown in Figure \ref{fig:dustcorrplot}. Future work could improve upon our results by modeling and removing dust and synchrotron emission from the LRGs.

Our $\bar{y}$ estimates include a contribution from nearby halos, known as the two-halo term (e.g. \cite{Hill2018}), which biases them high when assuming we have only one halo per source. Compared to the \cite{2016arXiv160702442B} simulations, our sample includes lower mass halos and our measurements include a beam, so they will have a larger two-halo contribution. We estimate this contribution using \texttt{Mop-c GT} (see Appendix A of \cite{Amodeo20} for details on the implementation) and we find that our $\bar{y}$ values are biased high by a factor between 2\% and 10\% in the luminosity bins of interest. The two-halo bias does not significantly affect the results presented in this work, but will need to be accounted for in future higher signal-to-noise analyses. 

The $\bar{y}$-$\bar{\tau}$ relationship (Equation \ref{eqn:ysim}) also carries associated systematics. These uncertainties were estimated based on differences between radiative cooling and AGN feedback sub-grid physics models in Battaglia (2017) \cite{2016arXiv160702442B}. While this does not encompass the wide variety of existing sub-grid models for the ICM, the two models are very different, and contrast in their inclusion of AGN feedback \cite{2016arXiv160702442B}. In addition to the quoted systematic uncertainty between the different models, we conjectured in DB17 that there is also uncertainty in extrapolating from the larger masses in the hydrodynamical simulations to the lower mass objects in the DR11 sample used in DB17 (Section \ref{sec:taucompare}). New hydrodynamical simulations of lower mass clusters and groups would help address this concern.

\subsection {kSZ signal extraction}
\label{sec:kSZsignals}

As described in C21, the mean pairwise momentum of groups and clusters as a function of their comoving separation distance is negative at and around separations of 25-50 Mpc, implying that they are moving towards one another on average due to gravity \cite{1991ApJ...374L...1H,1994MNRAS.271..976N,Juszkiewicz:1998xf,Sheth:2000ff}. While the 3-dimensional momentum of the groups is not easily measurable, the mean pairwise momentum $p$ can be still estimated from the line-of-sight component of the momenta \cite{Ferreira:1998id}. The kSZ signal of a given cluster is directly proportional to this line-of-sight momentum: $\delta T_{\rm kSZ,i} \propto -\mathbf{p}_i\cdot\mathbf{r}_i$, where the unstated multiplicative factors depend on the properties of the cluster (density profile, including angular extent in the sky) and on the pixel scale and angular resolution of the CMB experiment. 

The significance of the kSZ measurement is determined by a fit to the analytic prediction of linear perturbation theory for the pairwise velocity \cite{1991ApJ...374L...1H,1994MNRAS.271..976N,Juszkiewicz:1998xf,Sheth:2000ff,Bhattacharya:2007sk,Mueller:2014dba,Mueller:2014nsa}, rescaled by the factor $-\bar{\tau}T_{\rm CMB}/c$, where $\bar{\tau}$, the average halo optical depth of the galaxy sample used for the pairwise momentum estimator, is the free parameter of the fit.

\begin{table*}
\begin{center}
\setlength\extrarowheight{1.5pt}
\begin{tabular}{c|c|c|c|c|c|}
\cline{2-6}
\multicolumn{1}{c|}{\multirow{2}{*}{}} & \multicolumn{2}{c|}{{\seof}} &
    \multicolumn{2}{c|}{{\sen}} & \multicolumn{1}{c|}{{\ILC}}\\
\cline{2-6}
Bin & $\delta \rm{T}_{\rm{tSZ, corr.}} (\mu$K) & $\bar{y}/10^{-7}$ & $\delta \mathrm{T}_{\mathrm{tSZ, corr.}} (\mu$K) & $\bar{y}/10^{-7}$ & $\bar{y}/10^{-7}$  \\
\hline
L43     &   -0.14 $\pm$ 0.02  & 0.53 $\pm$ 0.09  &  -0.21 $\pm$ 0.02 & 0.51 $\pm$ 0.06 & 0.54 $\pm$ 0.08\\
L61     &   -0.21 $\pm$ 0.03  & 0.79 $\pm$ 0.11  &  -0.29 $\pm$ 0.03 & 0.70 $\pm$ 0.07 & 0.78 $\pm$ 0.10\\
L79$^+$ &   -0.32 $\pm$ 0.03  & 1.22 $\pm$ 0.13  &  -0.47 $\pm$ 0.04 & 1.12 $\pm$ 0.11 & 1.28 $\pm$ 0.14\\
L98     &   -0.57 $\pm$ 0.06  & 2.18 $\pm$ 0.22  &  -0.77 $\pm$ 0.06 & 1.84 $\pm$ 0.15 & 2.19 $\pm$ 0.22 \\
L116    &   -0.89 $\pm$ 0.10  & 3.42 $\pm$ 0.37  &  -1.19 $\pm$ 0.10 & 2.86 $\pm$ 0.24 & 3.52 $\pm$ 0.34 \\
\hline
L43D$^+$ &  -0.03 $\pm$ 0.03   & 0.11 $\pm$ 0.13 & -0.08 $\pm$ 0.04  & 0.20 $\pm$ 0.09  & 0.13 $\pm$ 0.11\\
L61D$^+$ &  -0.10 $\pm$ 0.05   & 0.39 $\pm$ 0.18 & -0.13 $\pm$ 0.05  & 0.31 $\pm$ 0.11  & 0.28 $\pm$ 0.12\\
L79D     &  -0.11 $\pm$ 0.05   & 0.41 $\pm$ 0.18 & -0.22 $\pm$ 0.05  & 0.53 $\pm$ 0.13  & 0.46 $\pm$ 0.16 \\
L98D     &  -0.25 $\pm$ 0.07   & 0.97 $\pm$ 0.27 & -0.35 $\pm$ 0.08  & 0.85 $\pm$ 0.19  & 0.78 $\pm$ 0.28 \\

\hline
\end{tabular}
\caption{Thermal SZ results from the {\seof}, {\sen}, and {\ILC} map analyses, along with 1$\sigma$ jackknife uncertainty estimates. Dust-corrected stacked tSZ signals $\delta T_{\rm tSZ}$ and $\bar{y}$ are given for the two coadded temperature maps, and $\bar{y}$ for the {\ILC} Compton-$y$ map. For the DR5 f150 map, the \textit{Herschel} dust correction is applied, and the uncertainties associated with these corrections are propagated into the cited jackknife uncertainties. For the DR5 f090 map, the \textit{Herschel} dust correction and the f090 beam correction scaling factor are applied. The $\bar{y}$ results from the disjoint bins shared with C21 (marked as $^+$bins) are shown in Figure \ref{fig:simplot}.}\label{ytable}
\end{center}
\end{table*}

\subsection{Theoretical $\bar{\tau}$ estimate}
\label{sec:theorytau}

For comparison with our measurements, we calculate theoretical estimates for the mean optical depths $\tauth$ for each of the luminosity bins. We follow the derivation in \citet{2016arXiv160702442B} using an NFW profile to estimate the optical depth in a given aperture,
\begin{equation}
    \tauth = \sigt x_\rmn{e} X_\rmn{H} (1 - f_\star) f_\rmn{b} \frac{M_\rmn{vir}(<\theta_{2.1^\prime})}{d_A^2m_p}.
\label{eq:tauobs}
\end{equation}
Here $\sigt$ is the Thomson cross-section, $x_\rmn{e}$ is the electron fraction defined as $x_\rmn{e} = (X_\rmn{H} + 1) / (2X_\rmn{H})$, $X_\rmn{H}$ is the primordial hydrogen mass fraction ($X_\rmn{H} = 0.76$),  $f_\star$ is the stellar mass fraction of the halo, $f_\rmn{b}$ is the universal baryon fraction ($\Omega_\rmn{b}/\Omega_\rmn{M}$), $m_\rmn{p}$ is the proton mass, and $d_A$ is the angular diameter distance to mean redshift of our sample. The parameter value for $f_\rmn{b} = 0.157$ is set from the cosmological parameters we choose. The value for $f_\star$ is inferred from the stellar mass-halo mass relation from abundance matching as described in \cite{Kravtsov2014}. We define the parameter $f_\rmn{c}$ = $\bar{\tau}_{{\rmn obs}}/\tauth$ to compare the estimated $\bar{\tau}$ values (Table \ref{theorytable}) to the theoretically predicted values. This parameter represents the fraction of theoretically predicted optical depth obtained by the two SZ measurements, and is of interest to compare the consistency of the two optical depth estimate methods.

\begin{table*}
\begin{center}

\hspace*{-1.5cm}
\begin{tabular}{|c|c|c|c|c||c|c|}

\cline{3-7}
\multicolumn{2}{c}{\multirow{2}{*}{}} & \multicolumn{5}{|c|}{{\seof}} \\
\hline
{} & $\tauth$ & $\bar{\tau}_{\rm tSZ}$ & $\sigma_{\rm sys.}$  & {} & $\bar{\tau}_{\rm kSZ}$ [C21]&  {}\\
Bin & $[10^{-4}]$ & $[10^{-4}]$ & $[10^{-4}]$  & $f_\rmn{c, tSZ} \pm ({\rm stat., sys.})$ & $[10^{-4}]$ & $f_\rmn{c, kSZ}$ \\
\hline 
L43$^{**}$ & 1.39 & 1.28 $\pm$ 0.10 & 0.27 & 0.92 $\pm$ (0.07, 0.20) & 0.54 $\pm$ 0.09 & 0.39 $\pm$ 0.06 \\
L61        & 1.77 & 1.55 $\pm$ 0.11 & 0.30 & 0.88 $\pm$ (0.06, 0.17) & 0.69 $\pm$ 0.11 & 0.39 $\pm$ 0.06 \\
L79$^+$    & 2.42 & 1.92 $\pm$ 0.10 & 0.34 & 0.79 $\pm$ (0.04, 0.14) & 0.88 $\pm$ 0.18 & 0.36 $\pm$ 0.07 \\
L98        & 3.35 & 2.55 $\pm$ 0.12 & 0.39 & 0.76 $\pm$ (0.04, 0.12) & {}              & {}      \\
L116       & 4.44 & 3.18 $\pm$ 0.17 & 0.43 & 0.72 $\pm$ (0.04, 0.10) &  {}             & {}     \\
\hline
L43D$^+$   & 0.70 & 0.59 $\pm$ 0.35 & 0.17 & 0.85 $\pm$ (0.50, 0.24) & 0.46 $\pm$ 0.24 & 0.66 $\pm$ 0.34 \\
L61D$^+$   & 1.06 & 1.10 $\pm$ 0.25 & 0.25 & 1.04 $\pm$ (0.24, 0.23) & 0.72 $\pm$ 0.26 & 0.68 $\pm$ 0.25 \\
L79D       & 1.53 & 1.12 $\pm$ 0.24 & 0.25 & 0.74 $\pm$ (0.16, 0.16) & {}              & {}     \\
L98D       & 2.09 & 1.71 $\pm$ 0.23 & 0.33 & 0.82 $\pm$ (0.11, 0.16)  & {}              & {}     \\
\hline 
\multicolumn{2}{c}{\multirow{2}{*}{}} & \multicolumn{5}{|c|}{{\sen}} \\
\hline
{} & $\tauth$ & $\bar{\tau}_{\rm tSZ}$ & $\sigma_{\rm sys.}$  & {} & $\bar{\tau}_{\rm kSZ}$ [C21]&  {}\\
Bin & $[10^{-4}]$ & $[10^{-4}]$ & $[10^{-4}]$  & $f_\rmn{c, tSZ} \pm ({\rm stat., sys.})$ & $[10^{-4}]$ & $f_\rmn{c, kSZ}$ \\
\hline
L43$^{**}$ & 1.39 & 1.25 $\pm$ 0.07 & 0.27 & 0.89 $\pm$ (0.05, 0.19) & 0.65 $\pm$ 0.13 & 0.47 $\pm$ 0.09 \\
L61        & 1.77 & 1.46 $\pm$ 0.07 & 0.29 & 0.82 $\pm$ (0.04, 0.16) & 0.82 $\pm$ 0.17 & 0.46 $\pm$ 0.10 \\
L79$^+$    & 2.42 & 1.84 $\pm$ 0.08 & 0.32 & 0.76 $\pm$ (0.03, 0.13) & 0.79 $\pm$ 0.27 & 0.33 $\pm$ 0.11 \\
L98        & 3.35 & 2.35 $\pm$ 0.10 & 0.37 & 0.70 $\pm$ (0.03, 0.11) & {}              & {} \\
L116       & 4.44 & 2.91 $\pm$ 0.12 & 0.41 & 0.66 $\pm$ (0.03, 0.09) & {}              & {} \\
\hline
L43D$^+$   & 0.70 & 0.79 $\pm$ 0.18 & 0.20 & 1.14 $\pm$ (0.25, 0.29) & 0.83 $\pm$ 0.34 & 1.19 $\pm$ 0.49  \\
L61D$^+$   & 1.06 & 0.98 $\pm$ 0.18 & 0.23 & 0.92 $\pm$ (0.17, 0.22) & 1.07 $\pm$ 0.35 & 1.01 $\pm$ 0.33 \\
L79D       & 1.53 & 1.27 $\pm$ 0.15 & 0.27 & 0.83 $\pm$ (0.10, 0.18) & {}              & {} \\
L98D       & 2.09 & 1.60 $\pm$ 0.18 & 0.30 & 0.77 $\pm$ (0.09, 0.15) & {}              & {} \\
\hline
\multicolumn{2}{c}{\multirow{2}{*}{}} & \multicolumn{5}{|c|}{{\ILC}} \\
\hline
{} & $\tauth$ & $\bar{\tau}_{\rm tSZ}$ & $\sigma_{\rm sys.}$  & {} & $\bar{\tau}_{\rm kSZ}$ [C21]&  {}\\
Bin & $[10^{-4}]$ & $[10^{-4}]$ & $[10^{-4}]$  & $f_\rmn{c, tSZ} \pm ({\rm stat., sys.})$ & $[10^{-4}]$ & $f_\rmn{c, kSZ}$ \\
\hline
L43$^{**}$ & 1.39 & 1.29 $\pm$ 0.09 & 0.27 & 0.92 $\pm$ (0.07, 0.19)  & 0.47 $\pm$ 0.12 & 0.34 $\pm$ 0.09  \\
L61        & 1.77 & 1.54 $\pm$ 0.09 & 0.30 & 0.87 $\pm$ (0.05, 0.17)  & 0.74 $\pm$ 0.15 & 0.42 $\pm$ 0.08 \\
L79$^+$    & 2.42 & 1.96 $\pm$ 0.11 & 0.34 & 0.81 $\pm$ (0.04, 0.14)  & 0.78 $\pm$ 0.23 & 0.32 $\pm$ 0.10 \\
L98        & 3.35 & 2.55 $\pm$ 0.13 & 0.39 & 0.76 $\pm$ (0.04, 0.12)  & {}               & {}      \\
L116       & 4.44 & 3.22 $\pm$ 0.15 & 0.43 & 0.73 $\pm$ (0.03, 0.10)  &   {}             & {}      \\
\hline
L43D$^+$   & 0.70 & 0.64 $\pm$ 0.26 & 0.17 & 0.91 $\pm$ (0.37, 0.25)  & 0.18 $\pm$ 0.32 & 0.26 $\pm$ 0.46 \\
L61D$^+$   & 1.06 & 0.93 $\pm$ 0.20 & 0.22 & 0.88 $\pm$ (0.19, 0.21)  & 0.69 $\pm$ 0.34 & 0.65 $\pm$ 0.32\\
L79D       & 1.53 & 1.19 $\pm$ 0.20 & 0.26 & 0.78 $\pm$ (0.13, 0.17)  & {}               & {}      \\
L98D       & 2.09 & 1.54 $\pm$ 0.27 & 0.30 & 0.74 $\pm$ (0.13, 0.14)  &  {}              & {}      \\
\hline

\end{tabular}
\caption{Optical depth estimates from the tSZ effect via hydrodynamic simulations, 1$\sigma$ statistical and systematic uncertainties, and fraction of theoretical estimates for mean optical depths ($f_\rmn{c}$, Section \ref{sec:theorytau}) for each luminosity bin and analyzed map. Statistical uncertainties on tSZ estimated optical depth are propagated from the tSZ AP jackknife uncertainty estimates and \textit{Hershcel} dust corrections. Systematic uncertainties are estimated using the Monte Carlo method taking into account the estimated systematic uncertainties in the $\bar{y}$-$\bar{\tau}$ relationship from simulations (Equation \ref{eqn:ysim}). For example, the $\bar{\tau}$ fit from the {\seof} tSZ results for $L>4.30\times 10^{10} L_{\odot}$, $1.28 \times 10^{-4}$, divided by the theoretical $\bar{\tau}$ estimate of 1.39$\times 10^{-4}$, yields $f_\rmn{c}=0.92$ for that galaxy sample. Selected $\bar{\tau}$ estimates from the pairwise kSZ effect from C21 with bootstrap uncertainties are listed along with $f_\rmn{c}$ for comparison. The fractions for the full galaxy sample ($^{**}$bin) are shown in Figure \ref{fig:barplot}, and kSZ results from the three disjoint bins shared with C21 ($^+$bins) are shown in Figure \ref{fig:simplot}. Uncertainties on mass estimates from luminosity have not been propagated through to the $\tauth$ estimate, so $f_\rmn{c}$ is best used for comparison and the study of relative trends.}\label{theorytable}
\end{center}
\end{table*}

\begin{figure*}[ht!]
\begin{center}
\includegraphics[width=12.9cm]{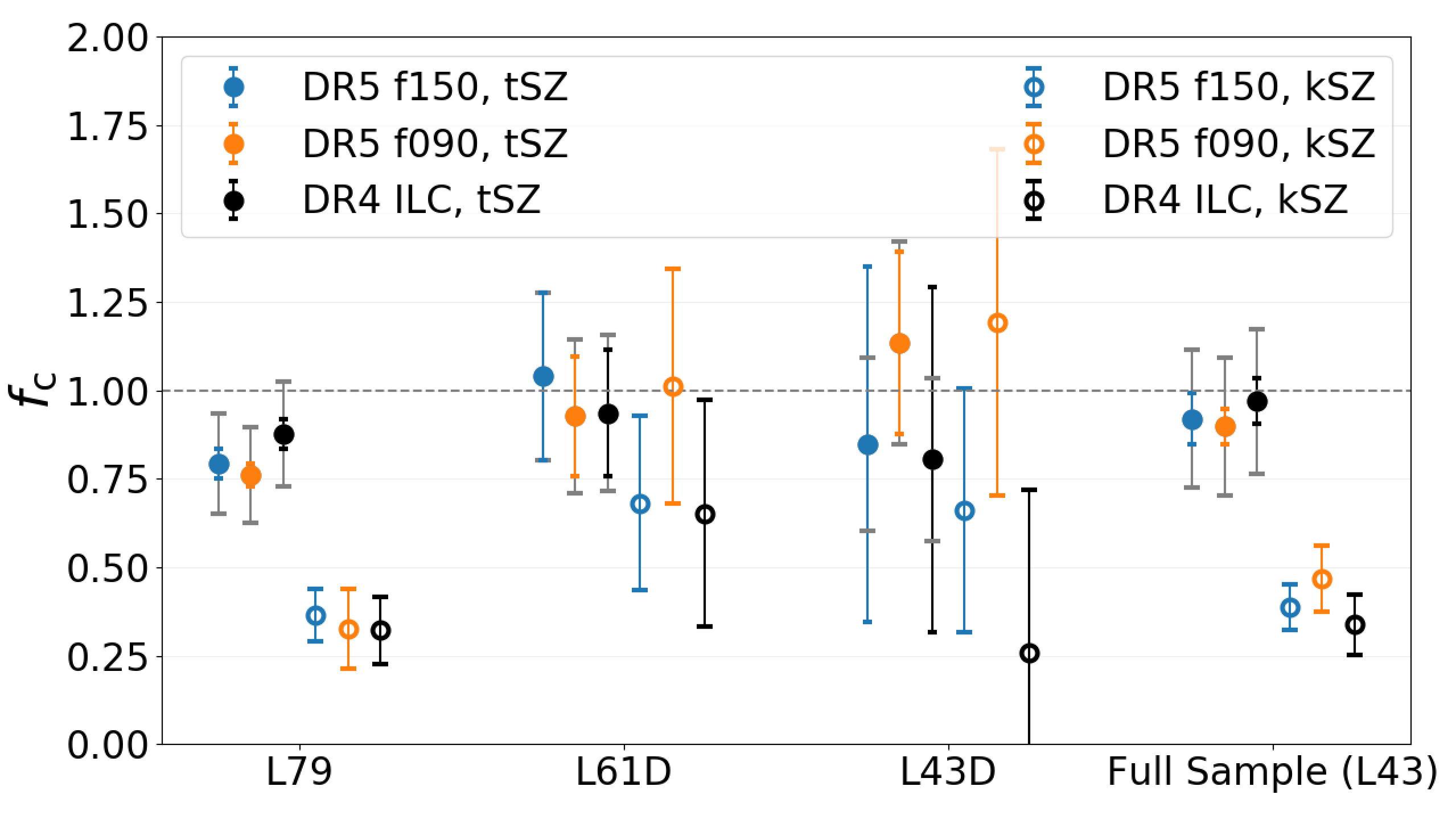}
\caption{Fraction of the theoretically predicted optical depth ($f_{\rm c}$) for the full DR15 sample. The estimates are extracted from tSZ measurements (filled circles) from three different maps ({\seof}: blue, {\sen}: orange, and ILC: black) and kSZ measurements from the same maps, as described in C21 (open circles). The tSZ measurements are converted to optical depth estimates using a scaling relationship from hydrodynamic simulations \cite{2016arXiv160702442B}. The tSZ jackknife uncertainties are plotted in color, and the systematic uncertainties from the simulation-based scaling relationship are plotted as grey bars. The plotted kSZ uncertainties are from bootstrap estimates. The kSZ and tSZ results agree within 1$\sigma$ in the L61D and L43D bins, while in the highest signal-to-noise L79 bin they differ at 2-3$\sigma$. This results in the 2-3$\sigma$ difference observed in the full sample, L43. The difference between the kSZ and tSZ results is discussed in Section \ref{sec:taucompare}. The kSZ results are lowest for the L79 bin.} 

\label{fig:barplot}
\end{center}
\end{figure*}

\begin{figure}[h!]
\begin{center}
\hspace*{-0.3cm}
\includegraphics[width=8.6cm]{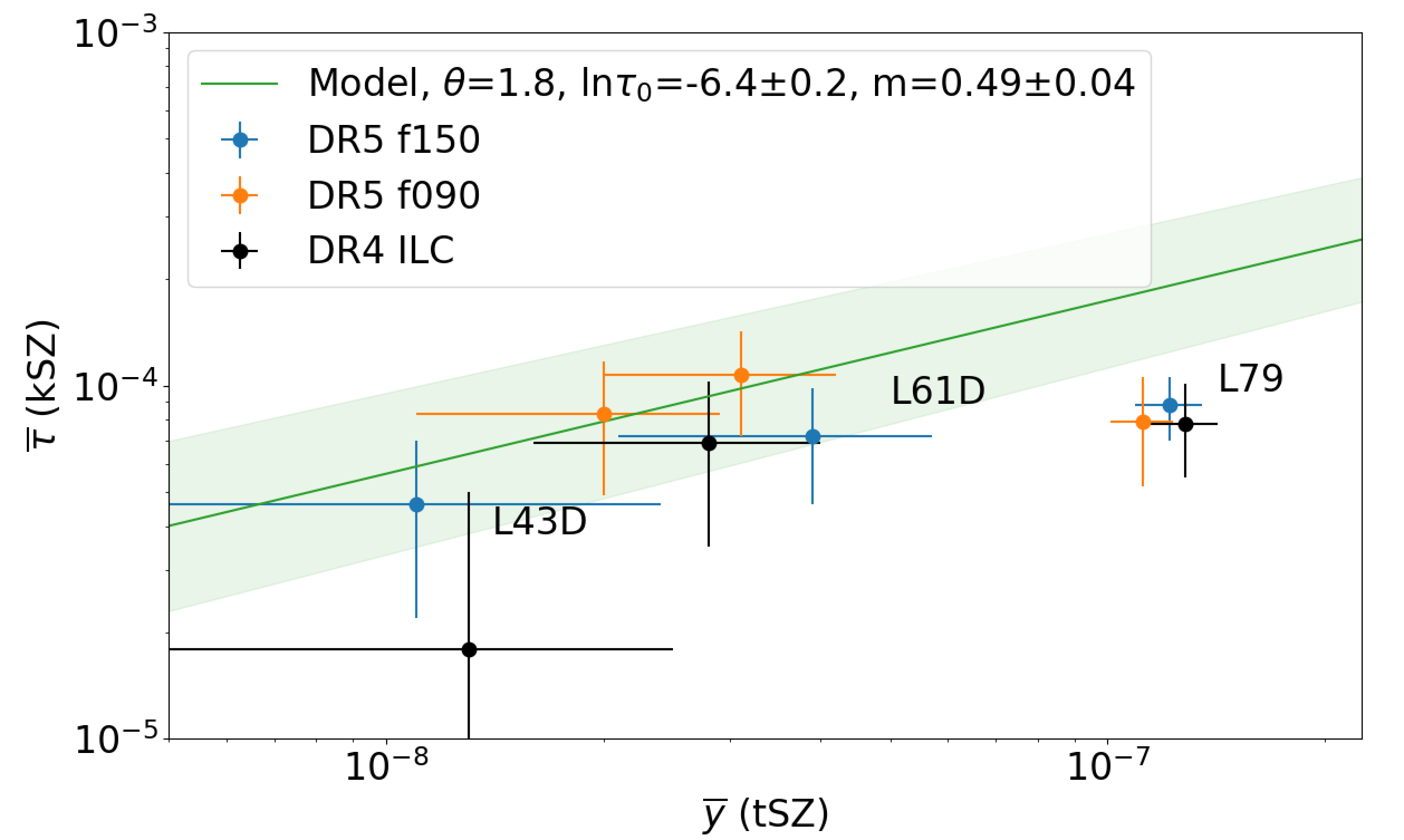}
\caption{Optical depth fits from kSZ signals versus average Compton-$y$ from the tSZ measurements for the three jointly analyzed disjoint luminosity bins with statistical error bars from jackknife estimates (bootstrap estimated uncertainties for the kSZ, see C21). The scaling relation between $\bar{\tau}$ and $\bar{y}$ from hydrodynamical simulations with aperture $\Theta=1.8^\prime$ (the closest scaling relation in \cite{2016arXiv160702442B} to our $2.1^\prime$ aperture) is plotted as the green model curve (Equation \ref{eqn:ysim}) with the 1$\sigma$ uncertainty envelope shaded. The tSZ and kSZ results in the L43D and L61D luminosity bins are consistent with the model, while the kSZ results in the L79 bin fall below the model line.}
\label{fig:simplot}
\end{center}
\end{figure}

\section{Results}\label{sec:results}
\subsection{tSZ measurements}\label{sec:tsz}

Table \ref{ytable} presents values of $\delta T_{\rm tSZ}$ for each luminosity bin along with the averaged Compton $\bar{y}$ parameter we calculated using the tSZ signal. These results are beam and dust corrected using our \textit{Herschel}-based estimates, and assume one source per filter (Section \ref{sec:sys}). The signal-to-noise ratios for the tSZ measurements are up to 10 for the {\seof} map, 12 for the {\sen} map, and 10 for the ILC map, with the highest signal-to-noise in the L98 bin for the coadded maps, and the L116 bin for the ILC map. Before the \textit{Herschel} dust corrections were applied and the uncertainties from those corrections propagated into the statistical tSZ uncertainty estimates, the highest signal-to-noise was seen in the highest luminosity bin (L116) for the coadded maps as well, which is as expected. The average Compton $\bar{y}$ from the tSZ signals in the three maps is presented in Figure \ref{fig:dtplots} along with signal-to-noise per disjoint bin. The $\bar{y}$ measurements are consistent across the maps we analyzed. 

The optical depths calculated using Eqn. \ref{eqn:ysim} range from 0.59$ \pm0.35 (\mathrm{stat.}) \pm0.17(\mathrm{sys.})\times10^{-4}$ to 3.22$ \pm0.15 (\mathrm{stat.}) \pm0.42 (\mathrm{sys.}) \times10^{-4}$ for all nine luminosity bins and three maps, as reported in Table \ref{theorytable}. Potential tSZ systematic effects are discussed in Section \ref{sec:sys}. 

\subsection{kSZ measurements}
\label{sec:tSZkSZcomp}

C21 presented a value of $\bar{\tau} = (0.69 \pm 0.11)\times 10^{-4}$ as the highest signal-to-noise best fit average optical depth.  It was derived from the mean pairwise momentum for the full sample of 343,647 sources and {\seof} map using bootstrap uncertainty estimates of the pairwise correlation covariance matrix.  In C21, we find that the analytical signal model is a good fit to the data with a best fit $\chi^2$ of 10 for 17 degrees of freedom. Selected best fit optical depths from the mean pairwise momentum fits and bootstrap estimated signal-to-noise are reported in Table \ref{theorytable}, and shown in Figures \ref{fig:barplot} and \ref{fig:simplot}, along with the results derived here.

\subsection{Optical depth comparisons}
\label{sec:taucompare}

Optical depth estimates from the two SZ effects are presented in Table \ref{theorytable} along with comparisons to theory. Figure {\ref{fig:barplot}} presents the fraction of theoretically predicted optical depth ($f_{\rm c}$) for the three independent joint analysis bins and the full sample, a representation of the baryon fraction we observe within the AP radius for those source samples. The tSZ and kSZ estimates are consistent within $1\sigma$ for the two lower mass disjoint bins (L61D and L43D), while they differ at 2-3$\sigma$ in the highest mass bin (L79), which is also the best constrained. This difference drives 2-3$\sigma$ differences between the tSZ and kSZ results in the cumulative bins. This difference may be decreased if the fixed cosmology assumed for the kSZ optical depth fits were allowed to vary. We look forward to comparing the two tracers in more detail in future work with even more sensitive data sets. The addition of more disjoint bins with sufficient numbers of sources for high signal-to-noise kSZ measurements will aid in the interpretation of any mass-dependent effects. Additionally, the halo mass input in the kSZ theory fits in C21 will continue to be investigated in future analyses. The comparisons with theoretical estimates of the optical depth suggest that between one third and all of the predicted baryons lie within the aperture size of radius 2.1$^\prime$, or 1.1 Mpc (for a mean redshift of z = 0.55) studied in this analysis. However, because uncertainties in the luminosity-mass relation for DR15 (Section \ref{sec:theorytau}) are difficult to accurately propagate through to $\tauth$, it is challenging to draw strong conclusions about baryon content based on our estimates of $f_{\rm c}$. For future high signal-to-noise measurements, careful treatment of the uncertainties, discussed in \cite{Kravtsov2014}, will be necessary for accurate interpretation of these SZ results. 
Here, the $f_{\rm c}$ values are used for making comparisons between the SZ estimates and studying trends. 

Figure \ref{fig:simplot} shows a comparison of $\bar{\tau}$ estimates from kSZ measurements and average Compton-$y$ from tSZ measurements to the power-law scaling relation using an AGN feedback model simulation relation (Equation \ref{eqn:ysim}) \cite{2016arXiv160702442B}. We expect significant covariance between the data from the different maps. Two of the three disjoint analysis bins are in agreement with the model, while the highest mass bin (L79) is an outlier. This appears to be driven by low kSZ $\bar{\tau}$ estimates for this bin (Fig. \ref{fig:barplot}).

The two-halo contribution to our tSZ measurements, discussed in Sec.~\ref{sec:sys}, does not affect these results. Propagating our $\bar{y}$ corrections to $\bar{\tau}_{tSZ}$ and then to $f_\rmn{c, tSZ}$ (filled circles in Fig. \ref{fig:simplot}), we get lower tSZ fractions by a factor of 4\% for the L43D bin and less for the other bins.

Our fits represent a step towards empirical $\bar{y}$-$\bar{\tau}$ relationships from a fit to kSZ and tSZ measurements. These measurements will thus serve as tests and checks for current and future cosmological simulations, and improve our understanding of galaxy formation and feedback models.

\section{Conclusion}\label{sec:conclusion}

We have presented estimates of halo optical depths from measurements of the tSZ effect made on the most recent multi-frequency ACT+\textit{Planck} maps in combination with LRG tracers from the SDSS BOSS DR15 catalog. By combining them with the kSZ measurements presented in C21, we compare estimates of optical depths from the two SZ effects and make progress towards empirical $\bar{y}-\bar{\tau}$ relationships from the SZ effects. We have improved our approach compared to our previous work, DB17, by analyzing the CMB map with the same AP filter for both the kSZ and tSZ analysis, and removing contaminating dust emission by estimating the contribution from dusty star-forming galaxies using \textit{Herschel} maps and a modified blackbody dust emission model.

The stacked tSZ signals were converted to an estimate of optical depth through a hydrodynamic simulation scaling model \cite{2016arXiv160702442B}, while the pairwise kSZ signals were fit to theoretical predictions to find best fit optical depths (C21). The two methods are independent of one another, and each of the SZ results is consistent over the three maps analyzed ({\seof}, {\sen}, and {\ILC}). The results from the tSZ and kSZ measurements agree with one another within 1$\sigma$ in the two lower mass disjoint bins, while they differ by 2-3$\sigma$ in the highest mass bin and thus the cumulative bin. Across all bins, the optical depth estimates from the SZ effects account for one third to all of the theoretically predicted baryon content. When comparing the tSZ $\bar{y}$ measurements and kSZ $\bar{\tau}$ results to the hydrodynamic model, two of the three bins analyzed are in agreement with the model.  

Using tSZ data appears to be a promising approach for obtaining accurate estimates of galaxy group and cluster optical depths and eventually estimating the mean pairwise velocity from pairwise momentum measurements, once a better understanding of the relationship between $\bar{y}$ and $\bar{\tau}$ is achieved. Our results help to move us from an era of measurement alone to one of interpretation based on the consistency of previously inaccessible quantities. 

With improved data from the complete ACT dataset \cite{Henderson:2015nzj} and current and upcoming projects such as CCAT-prime \cite{Stacey:2018}, the Simons Observatory \cite{SO:2019}, DESI \cite{DESI2019}, SPT-3G 
\cite{SPT3GBenson} and CMB-S4 \cite{Abazajian:2013oma}, we will achieve higher signal-to-noise measurements of the SZ signals and be able to probe the baryon content of galaxy clusters and groups and large-scale structure further. Improved multi-frequency data will enable precise measurements of optical depths and peculiar velocities simultaneously for large samples and for single sources, potentially sensitive enough to measure the missing baryons between groups. With these data, the SZ signals are anticipated to become valuable cosmological probes that are complementary to current observables. 

\section{Acknowledgements}
EMV acknowledges support from the NSF Graduate Research Fellowship Program under Grant No. DGE-1650441. MDN acknowledges support from NSF CAREER award 1454881. VC and RB acknowledge support from DoE grant DE-SC0011838, NASA ATP grant 80NSSC18K0695, NASA ROSES grant 12-EUCLID12-0004 and funding related to the Roman High Latitude Survey Science Investigation Team. NB acknowledges support from NSF grant AST-1910021, NASA ATP grant 80NSSC18K0695 and from the Research and Technology Development fund at the Jet Propulsion Laboratory through the project entitled ``Mapping the Baryonic Majority''. EC acknowledges support from the STFC Ernest Rutherford Fellowship ST/M004856/2 and STFC Consolidated Grant ST/S00033X/1, and from the European Research Council (ERC) under the European Union’s Horizon 2020 research and innovation programme (Grant agreement No. 849169). SKC acknowledges support from NSF award AST-2001866. JD is supported through NSF grant AST-1814971.  RD thanks CONICYT for grant BASAL CATA AFB-170002.  RH acknowledges funding from the CIFAR Azrieli Global Scholars program and the Alfred P. Sloan Foundation.  JPH acknowledges funding for SZ cluster studies from NSF grant number AST-1615657. KM acknowledges support from the National Research Foundation of South Africa. CS acknowledges support from the Agencia Nacional de Investigaci\'on y Desarrollo (ANID) through FONDECYT Iniciaci\'on grant no. 11191125. ZX is supported by the Gordon and Betty Moore Foundation. 
 
This work was supported by the U.S. National Science Foundation through awards AST-0408698, AST-0965625, and AST-1440226 for the ACT project, as well as awards PHY-0355328, PHY-0855887 and PHY-1214379. Funding was also provided by Princeton University, the University of Pennsylvania, and a Canada Foundation for Innovation (CFI) award to UBC. ACT operates in the Parque Astron\'omico Atacama in northern Chile under the auspices of the Comisi\'on Nacional de Investigaci\'on Cient\'ifica y Tecnol\'ogica de Chile (CONICYT). 

Canadian co-authors acknowledge support from the Natural Sciences and Engineering Research Council of Canada. The Dunlap Institute is funded through an endowment established by the David Dunlap family and the University of Toronto. 
  
Funding for the Sloan Digital Sky 
Survey IV has been provided by the 
Alfred P. Sloan Foundation, the U.S. 
Department of Energy Office of 
Science, and the Participating 
Institutions. 
SDSS-IV acknowledges support and 
resources from the Center for High 
Performance Computing  at the 
University of Utah. The SDSS 
website is www.sdss.org.
SDSS-IV is managed by the 
Astrophysical Research Consortium 
for the Participating Institutions 
of the SDSS Collaboration including 
the Brazilian Participation Group, 
the Carnegie Institution for Science, 
Carnegie Mellon University, Center for 
Astrophysics | Harvard \& 
Smithsonian, the Chilean Participation 
Group, the French Participation Group, 
Instituto de Astrof\'isica de 
Canarias, The Johns Hopkins 
University, Kavli Institute for the 
Physics and Mathematics of the 
Universe (IPMU) / University of 
Tokyo, the Korean Participation Group, 
Lawrence Berkeley National Laboratory, 
Leibniz Institut f\"ur Astrophysik 
Potsdam (AIP),  Max-Planck-Institut 
f\"ur Astronomie (MPIA Heidelberg), 
Max-Planck-Institut f\"ur 
Astrophysik (MPA Garching), 
Max-Planck-Institut f\"ur 
Extraterrestrische Physik (MPE), 
National Astronomical Observatories of 
China, New Mexico State University, 
New York University, University of 
Notre Dame, Observat\'ario 
Nacional / MCTI, The Ohio State 
University, Pennsylvania State 
University, Shanghai 
Astronomical Observatory, United 
Kingdom Participation Group, 
Universidad Nacional Aut\'onoma 
de M\'exico, University of Arizona, 
University of Colorado Boulder, 
University of Oxford, University of 
Portsmouth, University of Utah, 
University of Virginia, University 
of Washington, University of 
Wisconsin, Vanderbilt University, 
and Yale University.

\appendix


\onecolumngrid

\section{CasJobs SDSS Query}\label{sec:query}

The SDSS data used in this work and C21 was downloaded from the SDSS Catalog Archive Server (CAS) via the SkyServer website in May of 2019. All of the CMASS and LOWZ galaxies should be included in this catalog, in addition to more recently released eBOSS galaxies.  We use this approach in defining the catalog to utilize as many SDSS LRGs as possible. The following query returned 602,461 objects. The positions, redshifts, cmodel and Petrosian magnitudes, extinction corrections, and object IDs are recorded for the objects. An RA and DEC cut is applied to target the survey area overlapping with our maps, and the redshifts of the objects are selected to match the range in DB17. ``ZWARNING$\_$NOQSO=0" indicates that the automated redshift estimate is reliable for the source. Objects with ``zWarning=0" have no known redshift issues. Querying with ``sciencePrimary$>$0" selects the best available unique set of spectra for the objects. Bitmasks are applied to exclude SDSS target flags: ``SpecObjAll.TILE$>=$ 10324" excludes incorrectly targeted LOWZ galaxies in early BOSS data, and the i-band fiber magnitude is selected to be the same as in DB17.

\begin{lstlisting}[breaklines=true]
SELECT
SpecObjAll.ra, SpecObjAll.dec, SpecObjAll.z,
PhotoObjAll.cModelMag_u,PhotoObjAll.cModelMag_g,PhotoObjAll.cModelMag_r,PhotoObjAll.cModelMag_i,PhotoObjAll.cModelMag_z,PhotoObjAll.cModelMagErr_u,PhotoObjAll.cModelMagErr_g,
PhotoObjAll.cModelMagErr_r,PhotoObjAll.cModelMagErr_i,PhotoObjAll.cModelMagErr_z,
PhotoObjAll.petroMag_u,PhotoObjAll.petroMag_g,PhotoObjAll.petroMag_r,PhotoObjAll.petroMag_i,PhotoObjAll.petroMag_z,PhotoObjAll.petroMagErr_u,PhotoObjAll.petroMagErr_g,
PhotoObjAll.petroMagErr_r,PhotoObjAll.petroMagErr_i,PhotoObjAll.petroMagErr_z,PhotoObjAll.extinction_u,PhotoObjAll.extinction_g,PhotoObjAll.extinction_r,PhotoObjAll.extinction_i,
PhotoObjAll.extinction_z,SpecObjAll.bestObjID into DR15_actplanck_catalog_wbestObjID_PetrANDcModel_20200902_EMV from SpecObjAll, PhotoObjAll, Photoz
 WHERE
((SpecObjAll.bestObjID = PhotoObjAll.objID) and (SpecObjAll.bestObjID = Photoz.objID))
and
   ((((SpecObjAll.ra BETWEEN 142.0 AND 180.0) and (SpecObjAll.dec BETWEEN -8.3 AND 22.0)) or ((SpecObjAll.ra BETWEEN 0.0 AND 142.0) and (SpecObjAll.dec BETWEEN -61.5 AND 22.0)) 
     or ((SpecObjAll.ra BETWEEN 246.0 AND 360.0) and (SpecObjAll.dec BETWEEN -61.5 AND 22.0)) or ((SpecObjAll.ra BETWEEN 180.0 AND 246.0) and (SpecObjAll.dec BETWEEN -8.3 AND 22.0)))    
and (SpecObjAll.ZWARNING_NOQSO = 0) and
(SpecObjAll.zWarning = 0) AND
(SpecObjAll.sciencePrimary>0) and
(SpecObjAll.z>0.049) and
(SpecObjAll.z<0.8) and
((((SpecObjAll.BOSS_TARGET1 & 0x0000000000000001) != 0) and
(SpecObjAll.TILE>= 10324)) OR
(((SpecObjAll.BOSS_TARGET1 & 0x0000000000000002)!=0) and
  (PhotoObjAll.fiber2Mag_i<21.5))))
  

\end{lstlisting}

\section{Luminosity Binning}\label{sec:lumbinning}

As discussed in Section \ref{sec:sdssdata}, luminosity bins for the joint tSZ and kSZ analyses with C21 were chosen based on luminosity cuts from DB17 ($L>7.9\times10^{10} L_{\odot}$ and $L>6.1\times10^{10} L_{\odot}$) as well as one lower luminosity cut, ($L>4.3\times10^{10} L_{\odot}$). The three disjoint bins were selected to have roughly equal spacing, each with $\sim$100,000 galaxies after cutting for analysis with the DR5 maps (Table~\ref{sourcetable}). The luminosity cuts are plotted over a histogram of the SDSS samples in Figure 9.

\begin{figure*}[ht!]
\begin{center}
\includegraphics[width=12.9cm]{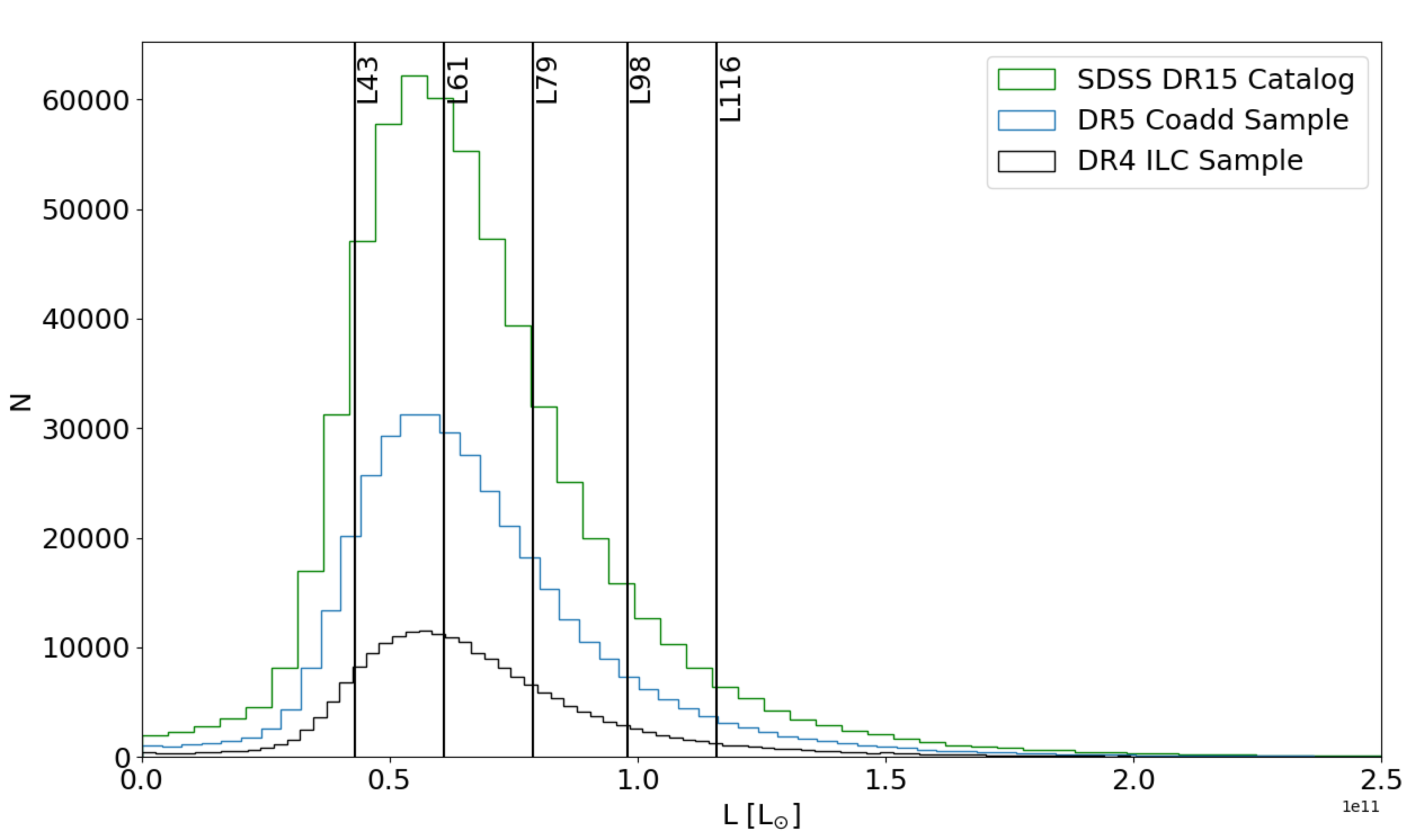}
\caption{Luminosity bin cuts for the SDSS DR15 catalog plotted over a histogram of the full sample (green), DR5 f150 and DR5 f090 selected analysis sample (blue), and DR4 ILC sample (black). The bottom three bins were selected to each have over 100,000 galaxies for the joint tSZ and kSZ analyses of the DR5 maps, while being roughly evenly spaced and overlapping with bin selection from the DB17 analysis. The top two bins were added for the tSZ analysis to study higher mass bins that have a strong tSZ signal, while also overlapping with DB17 bins.}
\end{center}
\label{fig:Figure9}
\end{figure*}

\bibliographystyle{apsrev}
\bibliography{Vavagiakis21}

\end{document}